\documentclass{article}
\usepackage[english]{babel}
\usepackage[letterpaper,top=2cm,bottom=2cm,left=3cm,right=3cm,marginparwidth=1.75cm]{geometry}
\usepackage{amsmath}
\usepackage{bm}
\usepackage{tabularx}
\newcolumntype{C}{>{\centering\arraybackslash}X} 

\usepackage{algorithm}
\usepackage{algpseudocode}
\algrenewcommand\algorithmicrequire{\textbf{Input:}}
\algrenewcommand\algorithmicensure{\textbf{Output:}}

\usepackage{graphicx}
\usepackage{subcaption}    
\usepackage[colorlinks=true, allcolors=blue]{hyperref}
\usepackage[percent]{overpic}

\title{Light scattering by random convex polyhedron in geometric optics approximation}
\author{Quan Mu \thanks{Corresponding author: mu.quan@foxmail.com} \\
Shenzhen MSU-BIT University, Beijing Institute of Technology}

\begin{document}
\begin{sloppypar}

\maketitle

\begin{abstract}
Based on the convex hull construction algorithm, a new geometrical model of ice crystals is proposed to investigate the scattering properties of cirrus clouds particles. Light scattering matrices involving complete polarization information are calculated in geometric optics approximation for randomly oriented large crystals with random and given convex polyhedron shape. The proposed model construction method and computational scheme of light scattering matrix works for any convex polyhedron within the scope of geometrical optics. To illustrate the broad applicability of the proposed ice crystal model, scattering matrices for three ice crystal examples with different geometrical shapes are calculated under a unified computational framework. Diffraction and absorption are not considered in this work. The calculated results for the classical hexagonal column model show overall agreement with those reported by other authors. The crystal model and scattering matrix computational framework developed in this study are applicable to radiative transfer simulations and remote sensing data interpretation in terrestrial and planetary atmospheres.
\end{abstract}

\section{Introduction}
Understanding micro-physical scattering properties of cirrus clouds crystals is fundamental to development of numerical radiative transfer model. Due to large varieties of crystal in morphology and size, numerical solutions to scattering characteristics by ice crystal is still a challenge in weather and climate research \cite{mishchenko2000light}. In the past four decades, several numerical methods have been developed, such as the finite-difference time domain (FDTD) method, the T-matrix method, the discrete dipole approximation (DDA) and geometric optics method (GOM) \cite{liou2016light}. In the late 20th century, researchers first explored the single scattering properties of simple ice crystal models like spheres, spheroids, and cylinders. Later, research extended to more complex models such as hexagonal prism, bullet rosette, hollow column, randomly shaped particles, as demonstrated in studies \cite{kokhanovsky1998dependence, macke1996applicability, takano1995radiative, GRYNKO2003319}. Based on the chemical foundation theory of ice crystal growth \cite{liou2016light, murray2015trigonal} and comparisons between mathematical modeling and physical remote sensing experimental results \cite{liou2016light, yang2018review}, the use of a hexagonal prism model has certain rationality. Calculations by many authors have shown that transitioning from idealized particle shapes to irregularly shaped particles leads to significant changes in their scattering properties. Therefore, many subsequent research efforts have attempted to construct irregularly shaped particle models. For example, work \cite{liu2014effective} describes a particle model constructed by randomly tilting the faces of a hexagonal prism according to a specified tilt distribution. In \cite{macke1996single}, a particle model based on fractal theory, more specifically, the Koch curve, was proposed. According to the constructed model, particles can be stretched or compressed in any direction \cite{liu2013effects}. In \cite{gasteiger2011modelling, shishko2019coherent}, To simulate randomly shaped particles, a rhombic bipyramidal crystal structure was used as the base geometry, which was then truncated by randomly oriented planes to generate ice crystals of arbitrary morphology. For the first time, we proposed using convex hulls as a random model of ice crystal particles and studied the scattering phase function of such particles \cite{mu2022computer, kargin2022numerical}.    

In this paper, we develop a new crystal model extending our previous work \cite{mu2022computer, kargin2022numerical}. Scattering matrices with complete polarization information of the proposed crystal model are computed based on Monte-Carlo method and ray tracing principle in the geometric optics (GO) regime. In section \ref{section 2} we first introduce the convex polyhedron construction method, then describe the coordinate system, ray tracing setup and scattering matrix computation scheme. Computational results and discussion of scattering matrices by random convex polyhedron and regular crystal are presented in section \ref{section 3}. Finally, conclusions and remarks are given in section \ref{section 4}. 

\section{Model description and computational scheme}
\label{section 2}
To compute light scattering matrices by randomly oriented crystals in geometric optics approximation, we first prepare a crystal model and introduce its construction method. Next, we describe ray tracing setup and coordinate system. Finally, we describe how to compute direction and polarization information of reflected and refracted rays. Based on those three stages, we have developed a program called \textit{Mueller Matrix of Convex Polyhedron (MMCP)}. To illustrate the computational procedure of the Mueller Matrix calculation for convex polyhedra, we provide a pseudocode representation of the algorithm (Algorithm \ref{alg:MMCP}). 

\begin{algorithm}
\caption{Mueller Matrix of Convex Polyhedron (\textit{MMCP})}
\label{alg:MMCP}
\begin{algorithmic}[1]
\Require Set of points $\mathcal{P}$
\Ensure Scattering matrix $M$

\State Construct convex polyhedron $ConvPoly$ from $\mathcal{P}$ using convex hull algorithm (Subsection \ref{subsection 2.1})

\For{$\alpha = 0$ to $\alpha_{max}$} \Comment{Euler angle 1}
    \For{$\beta = 0$ to $\beta_{max}$} \Comment{Euler angle 2}
        \For{$\gamma = 0$ to $\gamma_{max}$} (Eqs.(\ref{eq:rotate01}), (\ref{eq:rotate02})) \Comment{Euler angle 3}
            
            \For{photon $q = 0$ to $q_{max}$}
                \State Specify the incident ray of photon $q$ (Eqs. (\ref{eq1})–(\ref{eq4}))
                \State Trace the ray with reflection and refraction on $ConvPoly$ (Eqs.(\ref{eq:RT}), (\ref{Eq:Fresnel Coeff}))  
                \State \hspace{1.5em} until the photon exits the particle or the preset recursion depth is reached     
                \State Determine scattering angle $\theta_i, \phi_j$ of photon $q$ 
                \State Compute the $2\times2$ complex Jones matrix $J_{q}$ for the scattered photon (Eq. (\ref{Eq:Jp}))
                \State Convert the Jones matrix $J_{q}$ to a $4\times4$ real Mueller matrix $M_{q}$ (Eqs.(\ref{eq:convert J to M 01}), (\ref{eq:convert J to M 02}))   
                \State Accumulate $M_{q}$  into corresponding bins $[\theta_i, \phi_j]$
            \EndFor
            
        \EndFor
    \EndFor
\EndFor
\State \Return $M$
\end{algorithmic}
\end{algorithm}

The pseudocode summarizes the main steps, including the construction of the convex polyhedron from a set of 3D points, the looping over Euler angles to account for particle orientations, and the Monte Carlo simulation of photons. For each photon, the ray is traced through the polyhedron with reflection and refraction until it exits the particle or reaches a preset recursion depth, and the resulting Jones matrix is computed and converted into a Mueller matrix, which is accumulated into angular scattering bins. 

In the following subsections, we present these three parts in detail, along with additional computational aspects involved in the pseudocode.

\subsection{Convex polyhedron} 
\label{subsection 2.1}
The convex hull of a set of points is the minimal convex set that contains all the points. The problem of constructing convex hull of a finite set of points is a classical problem in computational geometry, with broad applications across many fields \cite{preparata2012computational}. Different algorithms and methods for computing convex hulls in two- or three-dimensional spaces have be studied extensively \cite{de2008computational}. In this paper, the problem of constructing convex three-dimensional bodies of arbitrary shapes is addressed in the context of numerical studies on the scattering properties of ice crystals in cirrus clouds. 

To compute convex hull of a set of given or randomly generated points, the incremental algorithm and the directed edges algorithm \cite{de2008computational,mu2022computer} are employed. The initial polyhedron is a tetrahedron whose vertices can be manually specified or generated random points distributed within a given volume according to a certain probability distribution. Each of the remaining points is processed sequentially. If a point is inside the current convex hull, no update is required. If it lies outside, it is added as a new vertex to form the updated convex hull. Details on the construction of three-dimensional convex bodies can be found in our previous work \cite{mu2022computer} or in books \cite{preparata2012computational, de2008computational}. 

\begin{figure}[htbp]
\centering
\includegraphics[width=0.3\linewidth]{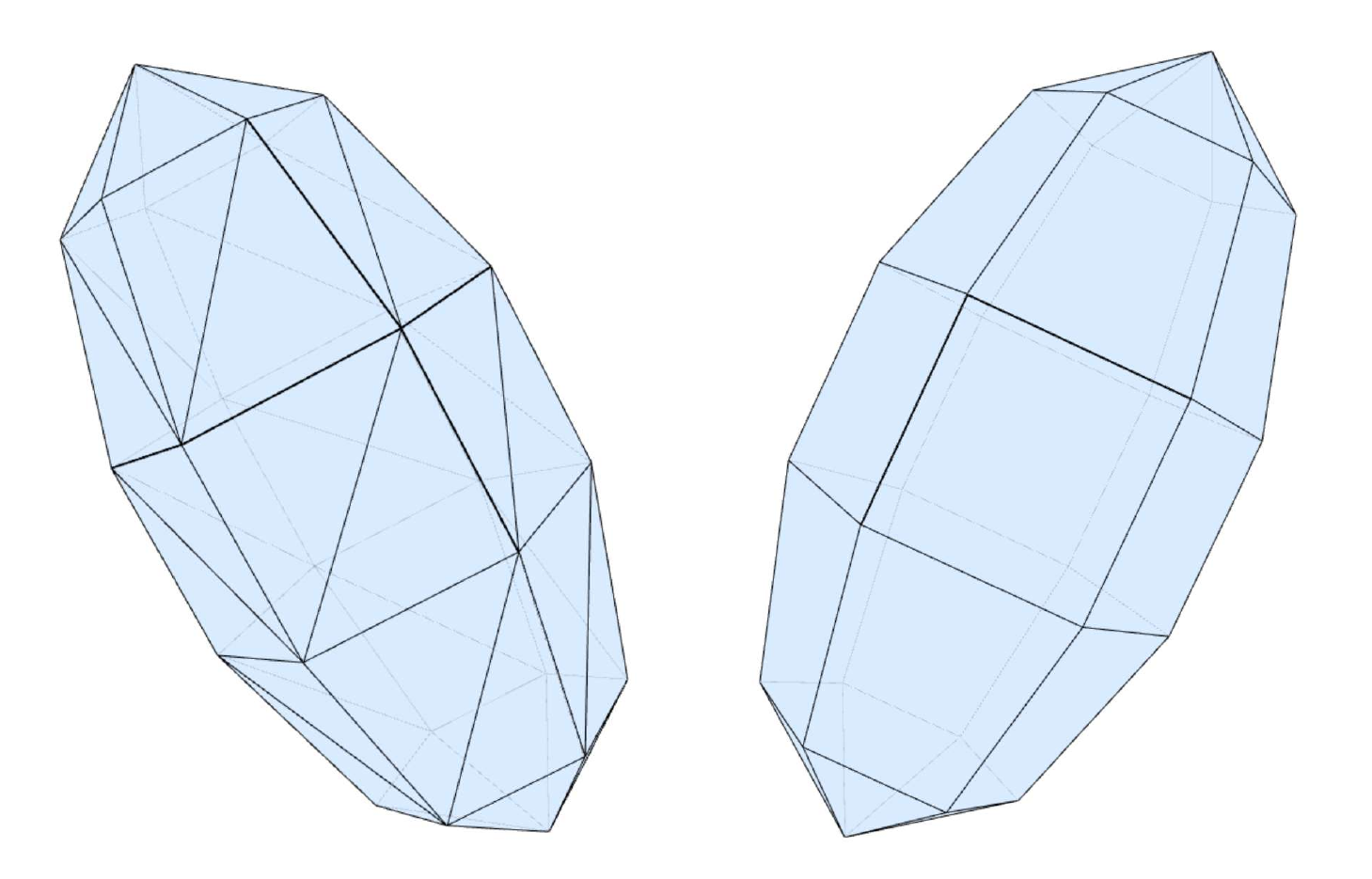}
\caption{\label{fig:3}Demonstration of mesh optimization: before (left) and after (right) merging coplanar triangular faces.}
\end{figure}

A key feature of our earlier convex hull model is that all its faces are triangular. In this paper, we present a more universal model by extending the convex hull model introduced in our previous works \cite{mu2022computer, kargin2024monte}. By detecting and merging coplanar triangular faces, the new convex polyhedron construction framework with optimized mesh can incorporate various commonly used mathematical models of regular ice crystal particles. Examples of such mesh optimization are demonstrated in Figure \ref{fig:3}. At the end of the convex polyhedron construction, useful information about the completed polyhedron is recorded and saved to files, for example, the number of faces, the number of vertices, and the coordinates of all vertices. Additionally, the vertices of each face are ordered such that the vector cross product of the edge vectors follows the right-hand rule, ensuring that the face normal points outward. This ordered vertex structure is particularly useful for calculating light scattering matrices using the ray tracing method. The code is written in C++ and includes functionality for visualizing polyhedron using the OpenGL library. The newly proposed convex polyhedron model in this study offers new methods and approaches for investigating problems associated with the vast geometric diversity of cirrus ice crystal particles. 

\begin{figure}[htbp]
\centering
\includegraphics[width=0.8\linewidth]{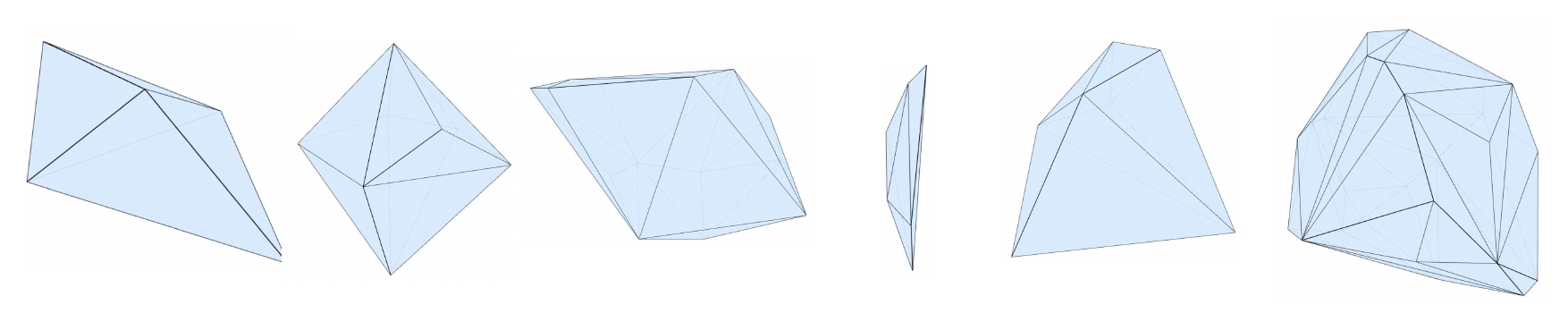}
\caption{\label{fig:1}Examples of random irregular convex polyhedron generated by the program \textit{MMCP}.}
\end{figure}

\begin{figure}[htbp]
\centering
\includegraphics[width=0.8\linewidth]{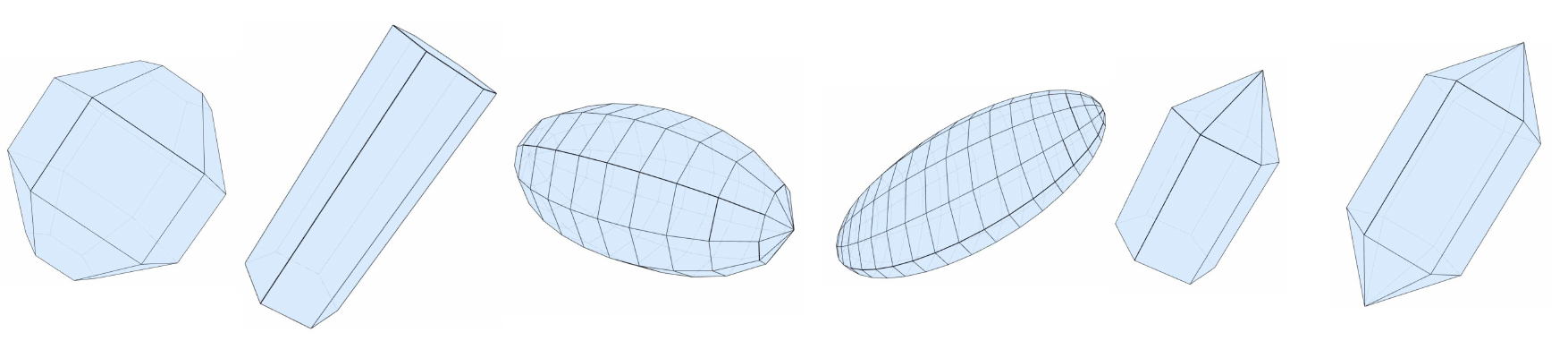}
\caption{\label{fig:2}Examples of regular convex polyhedron generated by the program \textit{MMCP}.}
\end{figure}

In this paper, we study two classes of particles. The first class is random irregular convex polyhedron, the second class is regular convex polyhedron, examples of corresponding shaped crystal are presented in Figure \ref{fig:1} and Figure \ref{fig:2}. 

\subsection{Ray tracing setup}
To specify an incident ray, a starting point $\bm{p}_0$ and a unit direction vector $\hat{\omega}_0$ should be defined. We first define a global Cartesian coordinate system \textit{OXYZ}, then the direction of an incident ray can be specified as 

\begin{equation}
\hat{\omega}_0=(\sin \theta \cos \varphi, \sin \theta \sin \varphi, \cos \theta), \label{eq1}
\end{equation}
where the polar angle $\theta \in [0, \pi]$ measures the angle from the positive \textit{Z} axis to $\hat{\omega}_0$, the azimuthal angle $\varphi \in [0, 2 \pi]$ is the signed angle measured from the the positive \textit{X} axis to the orthogonal projection of the unit direction vector $\hat{\omega}_0$ on the \textit{X-Y} plane. The angle $\varphi$ is defined positive if the rotation from positive \textit{X} axis is counterclockwise when viewed from the positive \textit{Z} axis.

After defining the direction of an incident ray, we now specify its starting point $\bm{p}_0$. Let $P$ be the plane with normal vector $\hat{\omega}_0$. To sample point $\bm{p}_0$ on the plane $P$, we define two additional unit vector $\hat{v}_0$ and $\hat{u}_0$ forming a right-handed orthonormal basis $(\hat{v}_0 , \hat{u}_0 , \hat{\omega}_0)$ , analogous to $(\hat{x} , \hat{y} , \hat{z})$ and satisfying the following relations:
\begin{equation}
    \hat{v}_0 \cdot \hat{u}_0=0,  \ \hat{v}_0 \times \hat{u}_0 =\hat{\omega}_0. \label{eq2}
\end{equation}
In practice, we first define a temporary vector $\bm{v}_t$ that is not parallel to $\hat{\omega}_0$. This can be either predefined or randomly generated using $(\sin \theta \cos \varphi, \sin \theta \sin \varphi, \cos \theta)$ with random $\theta, \varphi$. Then $\hat{u}_0$ and $\hat{v}_0$ are found by

\begin{equation}
     \hat{u}_0 =\frac{ \bm{v}_t \times \hat{\omega}_0}{\|\bm{v}_t \times \hat{\omega}_0\|}, \  \hat{v}_0=\hat{u}_0\times \hat{\omega}_0, \label{eq2.2.3}
\end{equation}
thus, the starting point $\bm{p}_0$ of an incident ray can be given as follows:

\begin{equation}
\bm{p}_0=\bm{p}_c+t \hat{u}_0+s \hat{v}_0, \label{eq3}
\end{equation}
where $\bm{p}_c$ is the center point of a circle $C$ on the plane $P$, $t$ and $s$ are random numbers sampled uniformly from interval $(-R_{max},R_{max})$, here $R_{max}=\max\{\|\bm{V}_i\|\}$ is the maximum distance among the convex polyhedron vertices $\{\bm{V}_i\}, i=1,...,N $. For practical purposes, $\bm{p}_c$ is set to

\begin{equation}
   \bm{p}_c=-2R_{max} \hat{\omega}_0. \label{eq4}
\end{equation}
It should be noted that, in this study, we assume that the geometric center of the ice crystal particle model coincides with the origin $O$ of the coordinate system \textit{OXYZ}, the value of $R_{{max}}$ is chosen such that the orthogonal projection of the ice crystal model onto the plane $P$ is entirely enclosed within the region of the circle $C$.

To sample the orientation of crystal particles, we introduce two schemes: in the fixed crystal, rotating ray (FCRR) approach, the crystal remains stationary while the incident ray is rotated; in the fixed ray, rotating crystal (FRRC) approach, the incident ray is fixed while the crystal is rotated. In the FCRR mode, to compute scattering matrix for randomly oriented crystal, the incident direction $\hat{\omega}_0$ is determined by a simulated unit vector with an isotropic distribution over the unit sphere, and the starting point $\bm{p}_0$ is sampled within the circle on the plane perpendicular to the incident direction using the accept–reject technique, a method commonly used in Monte Carlo simulations. In the FRRC mode, the orientation of a crystal is expressed by the Euler angles $(\alpha, \beta, \gamma)$. Specifically, the coordinates of crystal vertices $\{\bm{V}^{\prime}_i=(x^\prime_i,y^\prime_i,z^\prime_i), i=1,...,N \}$ after rotating can be obtained by a coordinate transformation in the form

\begin{equation}
    {\bm{V}^{\prime}_i}^T=R\ {\bm{V}_i}^T, \label{eq:rotate01}
\end{equation}
where the matrix $R$ is a rotation matrix that represents a composition of elemental rotations $R_Z(\gamma), R_Y(\beta), R_Z(\alpha)$ and it is given by
\begin{equation}
   R=R_Z(\gamma)\ R_Y(\beta)\ R_Z(\alpha)= 
   \begin{bmatrix}
       \cos \gamma  & -\sin \gamma  & 0\\
       \sin \gamma  & \cos \gamma   & 0\\
       0            &     0         & 1
   \end{bmatrix}
   \begin{bmatrix}
       \cos \beta    & 0  & \sin \beta\\
        0            & 1  & 0         \\
        -\sin \beta  & 0  & \cos \beta
   \end{bmatrix}
   \begin{bmatrix}
       \cos \alpha  & -\sin \alpha  & 0\\
       \sin \alpha  & \cos  \alpha   & 0\\
       0            &     0         & 1
   \end{bmatrix}.   \label{eq:rotate02}
\end{equation}
Note that in the FRRC approach, the crystal orientation is specified by performing three successive extrinsic rotations: a rotation about the fixed $Z$-axis by an angle $\alpha$, followed by a rotation about the fixed $Y$-axis by an angle $\beta$, and finally a rotation about the fixed $Z$-axis by an angle $\gamma$.

It should be noted that the FCRR mode is generally more computationally efficient than the FRRC mode for problems involving randomly oriented crystals, as the latter requires rotating all vertices for each orientation. However, for controlling crystal orientation, the FRRC mode is more straightforward, particularly when the study focuses on crystals with non-random orientations. 

In this study, we apply the hit-and-miss Monte Carlo method to trace photons in convex ice crystals \cite{liou2016light}. To determine whether an incident ray can intersect with a convex polyhedron, we employ the method proposed in \cite{Zhang:2004}. The technical details is well described in \cite{Zhang:2004} and is not repeated here. 

\subsection{Scattering matrix}
\label{subsection 2.3}
To describe the computational method for the light scattering matrix of a convex polyhedron, we adopt the vector form following the notation introduced by  \cite{yang2006light}. Let $E^i_{0v}$ and $E^i_{0u}$ denote the components of incident electric field $\bm{E}^i_0$ along the $\hat{v}_{0}$ and $\hat{u}_{0}$ directions, then the incident polarization configuration can be specified via the following expression:
\begin{equation} \label{Eq:Ei0}
    \bm{E}^i_0=E^i_{0v}\hat{v}_{0}+E^i_{0u}\hat{u}_{0}.
\end{equation}
Similarly, the scattered electric field $\bm{E}^s$ can be expressed as follows:
\begin{equation}
    \bm{E}^s=E^s_{v}\hat{v}_{s}+E^s_{u}\hat{u}_{s}.
\end{equation}
The relation between the incident and scattered field components can be described through a scattering Jones matrix $J$ as
\begin{equation}
    \bm{E}^s=J \bm{E}^i_0,\\
\end{equation}
In general, for $p$-th-order transmitted rays $(p > 2)$, the $2\times2$ complex scattering matrix $J$ is obtained by multiplying the corresponding appropriate transformation matrices as follows:
\begin{equation} \label{Eq:Jp}
    J=\varGamma^s_{p} T_p \varGamma_{p} R_{p-1} \varGamma_{p-1}\cdots R_2 \varGamma_2 T_1 \varGamma_1 \varGamma^i_{p}.
\end{equation}
Here, $R, T$ are reflection and transmission matrices, respectively, and they are defined by
\begin{equation}\label{eq:RT}
    R=
    \begin{bmatrix}
        R_v & 0\\
        0  & R_u
    \end{bmatrix}, 
    T=
    \begin{bmatrix}
        T_v & 0\\
        0  & T_u
    \end{bmatrix}. 
\end{equation} 
$\varGamma$ denotes a 2-D rotational matrix determined by direction cosines, which are listed in Table \ref{table01}.

\begin{table}[htbp]
\centering
\caption{\label{table01}Direction cosines defining the rotation matrix $\varGamma$ from the basis $(\hat{v}_{i},\hat{u}_{i})$ to new basis $(\hat{v}_{i+1},\hat{u}_{i+1})$}
\begin{tabularx}{0.8\textwidth}{C C C}
\hline
              & $\hat{v}_{i+1}$                     & $\hat{u}_{i+1}$  \\\hline
$\hat{v}_{i}$ & $\hat{v}_{i} \cdot \hat{v}_{i+1} $  & $\hat{v}_{i} \cdot \hat{u}_{i+1}$ \\
$\hat{u}_{i}$ & $\hat{u}_{i} \cdot \hat{v}_{i+1} $  & $\hat{u}_{i} \cdot \hat{u}_{i+1}$ \\

\hline
\end{tabularx}
\end{table}
The elements of matrices in Eq.(\ref{eq:RT}) are complex values given by the Fresnel coefficients \cite{born2019principles} as follows: 

\begin{equation} \label{Eq:Fresnel Coeff}
\begin{gathered}
  R_v=\frac{\cos{\theta_i}-m\cos{\theta}_t}{\cos{\theta_i}+m\cos{\theta}_t},
  R_u=\frac{m\cos{\theta_i}-\cos{\theta}_t}{m\cos{\theta_i}+\cos{\theta}_t},\\
  T_v=\frac{2 \cos{\theta_i}}{\cos{\theta_i}+m\cos{\theta}_t},
  T_u=\frac{2 \cos{\theta_i}}{m\cos{\theta_i}+\cos{\theta}_t}.  
\end{gathered}   
\end{equation}
Here, $\theta_i,\theta_t$ are the angle of incidence and angle of refraction, respectively, which are determined by the law of reflection and Snell’s law, and $m$ is the relative refractive index of an optical medium $2$ with respect to another reference medium $1$. In our work the $m$ is defined as
\begin{equation}
    m=\frac{n_2}{n_1},
\end{equation}
where $n_2$ is the refractive index of the scattering particle, $n_1$ is the the refractive index of air.   

\begin{figure}[htbp]
\centering
\includegraphics[width=0.8\linewidth]{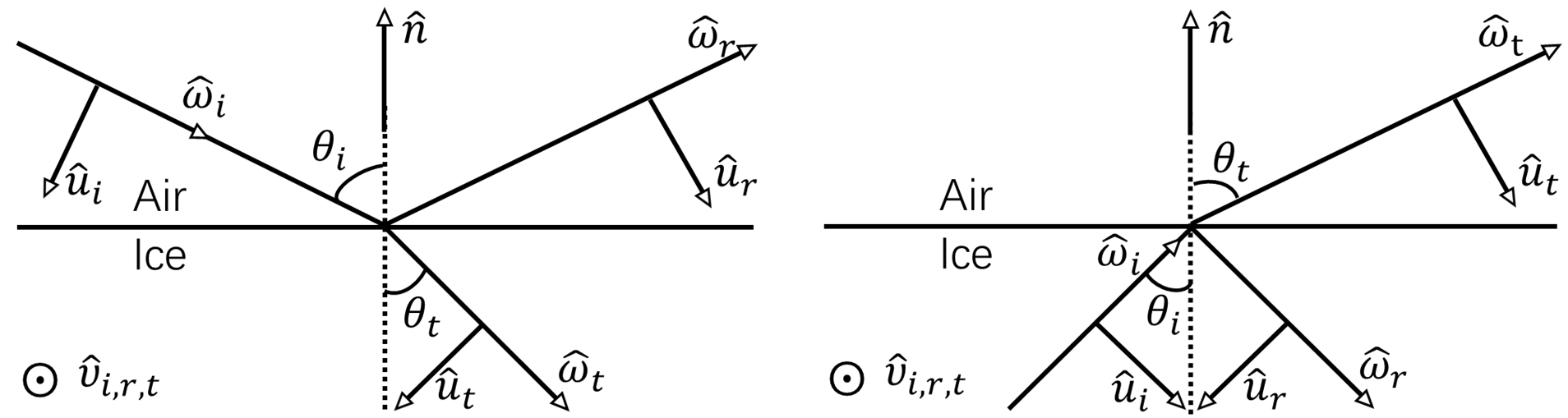}
\caption{\label{fig:4}Schematic representation of the incident, reflected, and refracted rays, together with the unit vectors defining the polarization configuration. The vectors $\hat{v}_{i,r,t}$ point out of the paper. Unlike Fig.2.2 in \cite{yang2006light}, the diagrams presented here are consistent with the assumption that all unit vectors $\hat{n}$, which are locally normal to the polyhedron faces, are directed outward.}
\end{figure}

As an example, for externally reflected ray, the Jones matrix is given by
\begin{equation}\label{Eq:J1}
    J=\varGamma^s_{1} R_{1} \varGamma_1 \varGamma^i_{1},
\end{equation}
where $\varGamma_1$ is necessary for applying the Fresnel formulas, as the $\bm{E}^i_0$ must be represented with respect to a new basis defined by $\hat{v}^i_1$ and $\hat{u}^i_1$ as follows:
\begin{equation}
      \hat{v}^i_{1}=\frac{ \hat{\omega}_0 \times \hat{n}_1}{\|\hat{\omega}_0 \times \hat{n}_1\|},
        \hat{u}^i_{1}=\hat{\omega}_r \times \hat{v}^i_{1},
\end{equation}
here $\hat{\omega}_r$ denotes the unit direction vector of reflected ray, and $\hat{n}_1$ represents the unit normal vector to the face of the polyhedron. In this study, the unit normal vector $\hat{n}$ of each face of the convex polyhedron is consistently defined to point outward from the particle, as illustrated in Figure \ref{fig:4}.

Thus, $\bm{E}^i_0$ need to be expressed in form:
\begin{equation} \label{Eq:Eii}
    \bm{E}^i_0=E^i_{1v}\hat{v}^i_{1}+E^i_{1u}\hat{u}^i_{1}.
\end{equation}
Then, the represented coordinates $E^i_{1v},E^i_{1u}$ can be specified as follows:
\begin{equation}
    \begin{bmatrix}
        E^i_{1v}\\
        E^i_{1u}
    \end{bmatrix}=\varGamma_{1}
    \begin{bmatrix}
        E^i_{0v}\\
        E^i_{0u}
    \end{bmatrix}, 
\end{equation}
where $\varGamma_{1}$ is a rotational matrix determined by direction cosines (see Table \ref{table01}) as follows:
\begin{equation}
    \varGamma_{1}=  
    \begin{bmatrix}
        \hat{v}_{0} \cdot \hat{v}_{1} & \hat{v}_{0} \cdot \hat{u}_{1}  \\
        \hat{u}_{0} \cdot \hat{v}_{1} & \hat{u}_{0} \cdot \hat{u}_{1}
    \end{bmatrix}, 
\end{equation}

Now, we can apply the Fresnel formulas to the electric fields associated with the incident ray, and the components of reflected field is given by
\begin{equation}
\begin{bmatrix}
        E^r_{1v}\\
        E^r_{1u}
    \end{bmatrix}=R_1
    \begin{bmatrix}
        E^i_{1v}\\
        E^i_{1u}
    \end{bmatrix}=R_1\varGamma_{1}
    \begin{bmatrix}
        E^i_{0v}\\
        E^i_{0u}
    \end{bmatrix}, 
\end{equation}
where $R_1$ is the reflection matrix defined in Eq.\eqref{eq:RT} and Eq.\eqref{Eq:Fresnel Coeff}. 
The matrix $\varGamma^s_{1}$ in Eq.\eqref{Eq:J1} is a rotation matrix that maps the polarization components from the reflected plane to the scattering plane. For externally reflected rays, it reduces to the identity matrix, since the reflected plane coincides with the scattering plane. In general, the matrix $\varGamma^s_{p}$ in Eq.~\eqref{Eq:Jp} for $p$-th-order transmitted rays $(p > 2)$ is a rotation matrix defined by the direction cosines.         

To obtain the scattering matrix, the incident field must be specified with respect to the directions parallel and perpendicular to the scattering plane. This requires applying the rotation matrix $\varGamma^i_{1}$ in Eq.\eqref{Eq:J1}. It should be noted, however, that since the scattering plane is not known in the beginning, this first rotation from the right, represented by $\varGamma^i_{1}$ in Eq.\eqref{Eq:J1}, or more generally $\varGamma^i_{p}$ in Eq.\eqref{Eq:Jp}, is in practice carried out as the final matrix multiplication. 

A more detailed description of the computational procedures of scattering matrix for large ice crystals can be found in work \cite{yang2006light}, where the ray-tracing technique has been thoroughly and systematically presented; therefore, it will not be repeated here.

To represent the solution in the form of a Mueller matrix, it is necessary to define Stokes parameters
\begin{equation}
    \bm{S}=(S_0,S_1,S_2,S_3 )=(I,Q,U,V).
\end{equation}
In this study, the Stokes parameters are defined as follows \cite{hulst1981light, hovenier2014transfer}:
\begin{equation}\label{Eq:Stokes}
\begin{aligned}
   I&=E_{v}E_{v}^*+E_{u}E_{u}^*,\\
   Q&=E_{v}E_{v}^*-E_{u}E_{u}^*,\\
   U&=E_{v}E_{u}^*+E_{u}E_{v}^*,\\
   V&= i(E_{v}E_{u}^*-E_{u}E_{v}^*).
\end{aligned}   
\end{equation}
Then, the corresponding \(4\times4\) Mueller matrix \(M \) is given by \cite{fujiwara2007spectroscopic}
\begin{equation}
    M= (M_{ij}(\hat{\omega}_0,\hat{\omega}))_{i,j=1}^{4}=\varGamma(J\otimes J^*)\varGamma^{-1}, 
    \label{eq:convert J to M 01}
\end{equation}
where $^*$ indicates the complex conjugate, and $\otimes$ is the Kronecker product. $\hat{\omega}_0$ is the unit direction vector of incident ray, and $\hat{\omega}$ is the unit direction vector of the scattering ray.  
\begin{equation}
    \varGamma=
    \begin{bmatrix}
        1& 0& 0& 1\\
        1& 0& 0& -1\\
        0& 1& 1& 0\\
        0& i& -i& 0\\
    \end{bmatrix}.\label{eq:convert J to M 02}
\end{equation}
 
It should be noted that the Stokes parameters defined in Eq.\eqref{Eq:Stokes} may differ in form from those adopted in other works, for instance, \cite{konoshonkin2015beam01}.

In summary, to obtain the Mueller matrix $M$, for each outgoing photon from the crystal, a $2\times2$ complex Jones matrix $J$ is first constructed by multiplying the appropriate rotation matrices and reflection or refraction matrix. This $2\times2$ complex matrix $J$ is then converted to a $4\times4$ real matrix. Subsequently, the $4\times4$ real matrix is summed up into a corresponding angular bin. Finally, the Mueller matrix $M$ is normalized so that the first matrix element $M_{11}$ (i.e., the phase function) satisfies the following normalization condition:
\begin{equation} \label{Eq:normalization}
    \int_{\Omega}M_{11}(\hat{\omega}_0,\hat{\omega}) \mathrm{d}\hat{\omega}=1.
\end{equation}

Furthermore, the present computational framework provides the capability to selectively control the number of refractions and internal reflections, thereby facilitating a detailed analysis of the roles of individual or collective light paths in shaping particular features of the scattering patterns - for example, the formation of halos in cirrus clouds and rainbows in water clouds. Comparable studies can be found, for example, in works \cite{konoshonkin2015beam02, GRYNKO2003319}. The main goal of this study is to develop new particle geometries that provide a unified representation of particle shape construction and can be conveniently applied to the computation of scattering characteristics. In particular, the proposed geometrical models are designed to cover as many morphological possibilities as possible, enabling the investigation of how different particle shapes influence scattering characteristics. It should be noted that diffraction and absorption are not considered in our this study.

\section{Results and discussion}
\label{section 3}
To validate the proposed crystal model introduced in the previous section and the applied computational scheme, the computation of Mueller matrix for randomly oriented hexagonal column is first carried out. To demonstrate the broad applicability of the ice crystal particle model developed in this study, we further computed the scattering matrices for regular polyhedra as well as for randomly irregular particles. The corresponding results and discussions are provided in the following subsections. 

\subsection{Hexagonal column}
\begin{figure}[htbp]
    \centering  
    \begin{subfigure}{0.45\textwidth}
        \centering
        \begin{overpic}[width=\linewidth]{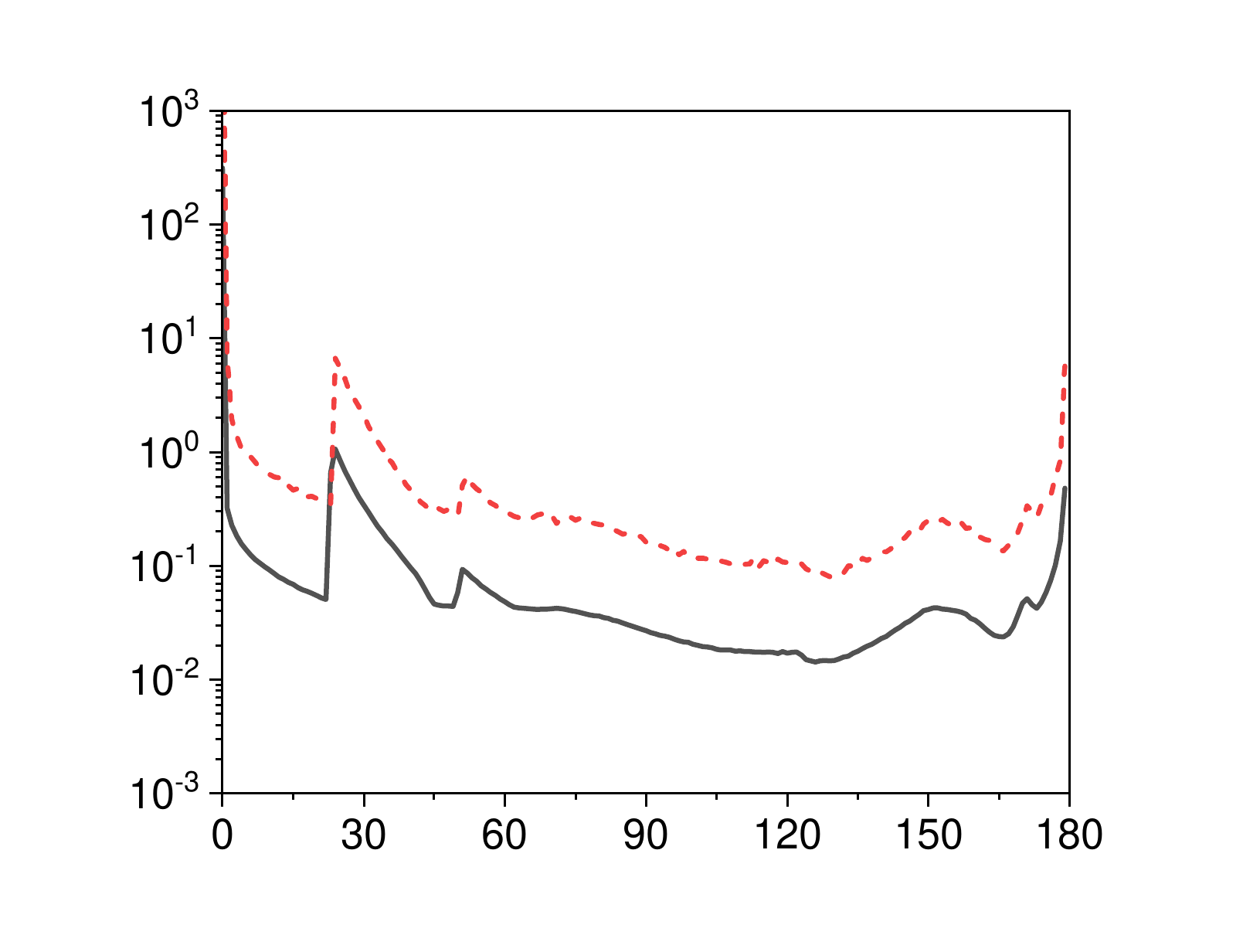}
            \put(60,40){\includegraphics[width=1.1cm]{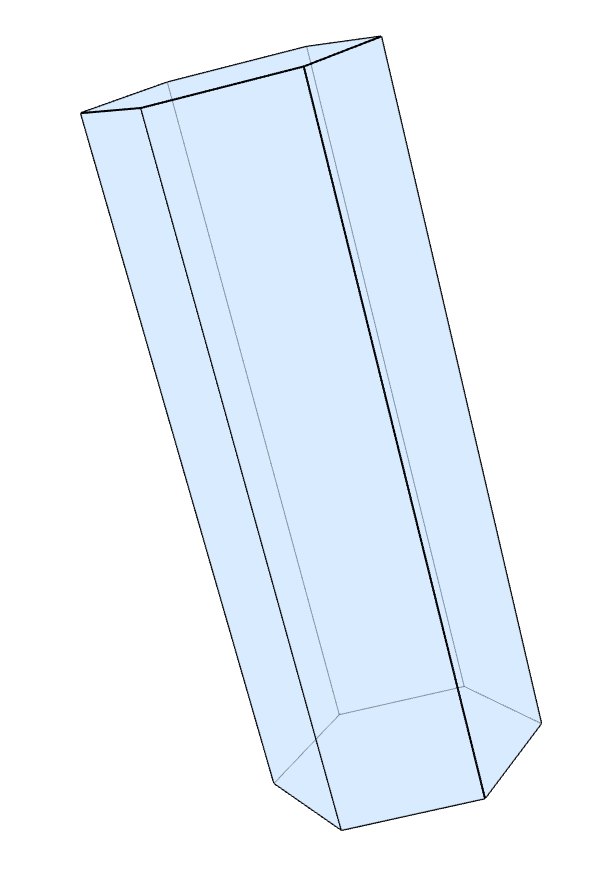}}
        \end{overpic}
        \caption{$M_{11}$}
        \label{subFig:5.a}
    \end{subfigure}
    \begin{subfigure}{0.45\textwidth}
        \centering
        \includegraphics[width=\linewidth]{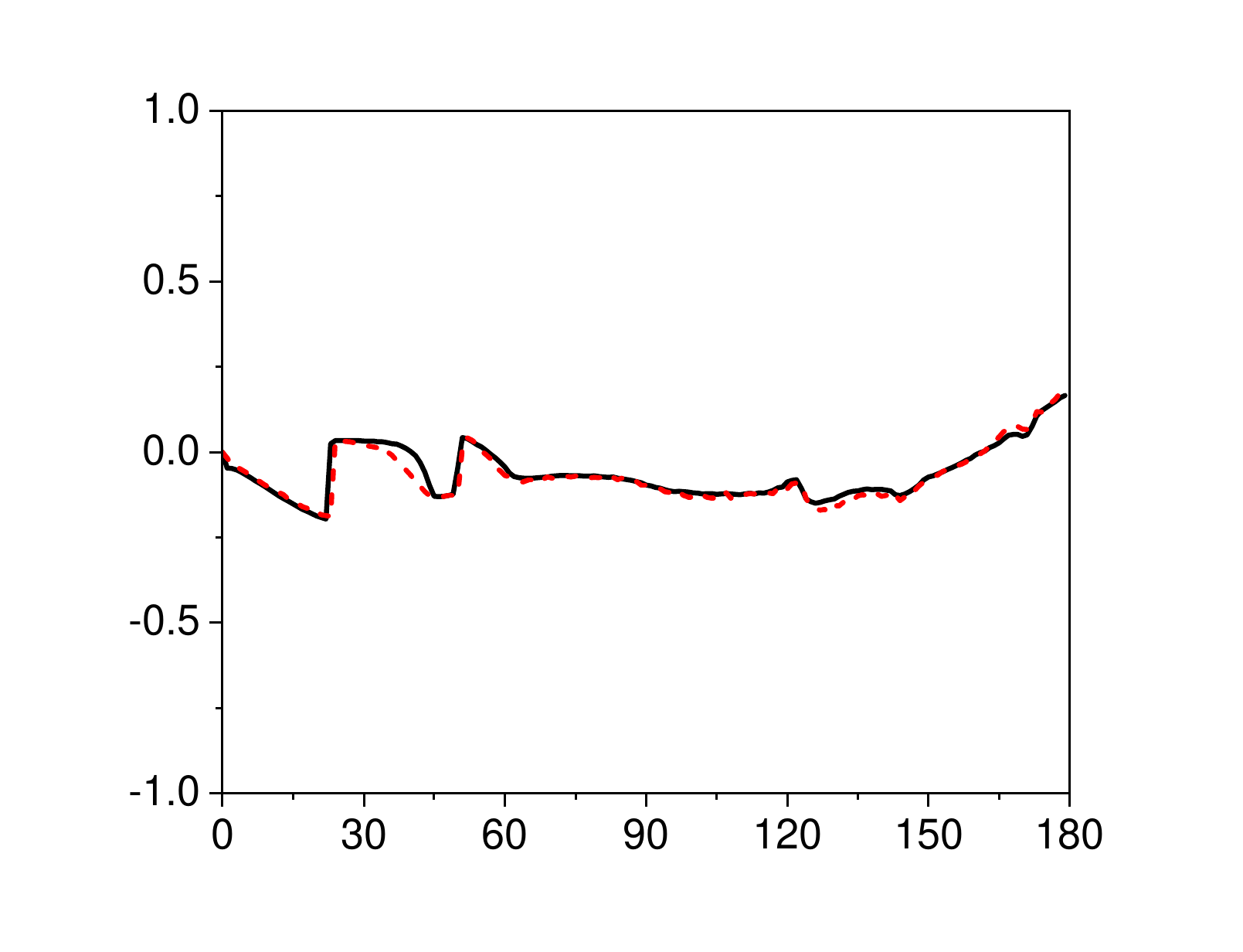}
        \caption{$M_{12}/M_{11}$}
         \label{subFig:5.b}
    \end{subfigure}      
    
     \begin{subfigure}{0.45\textwidth}
        \centering
        \includegraphics[width=\linewidth]{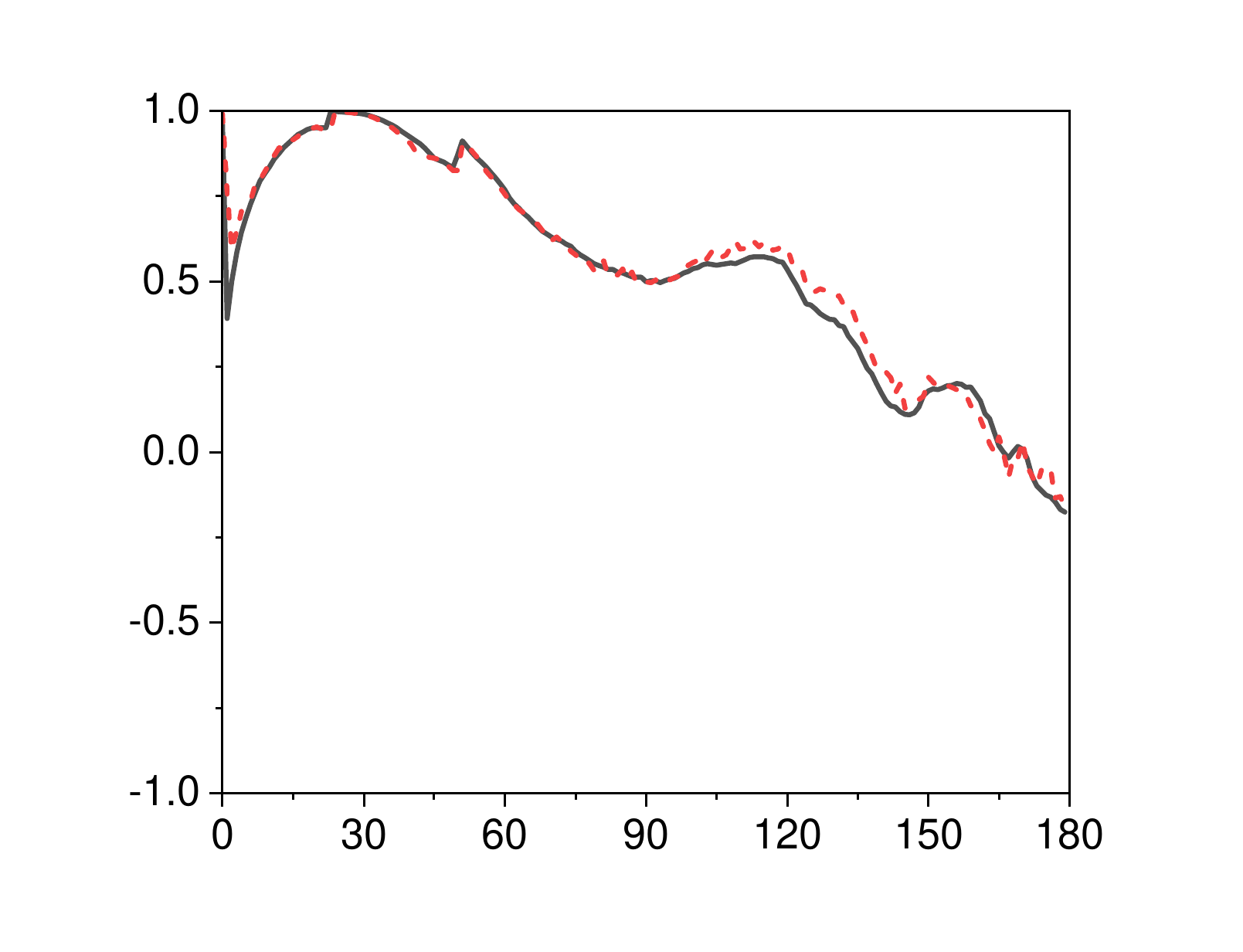}
        \caption{$M_{22}/M_{11}$}
         \label{subFig:5.c}
    \end{subfigure}
    \begin{subfigure}{0.45\textwidth}
        \centering
        \includegraphics[width=\linewidth]{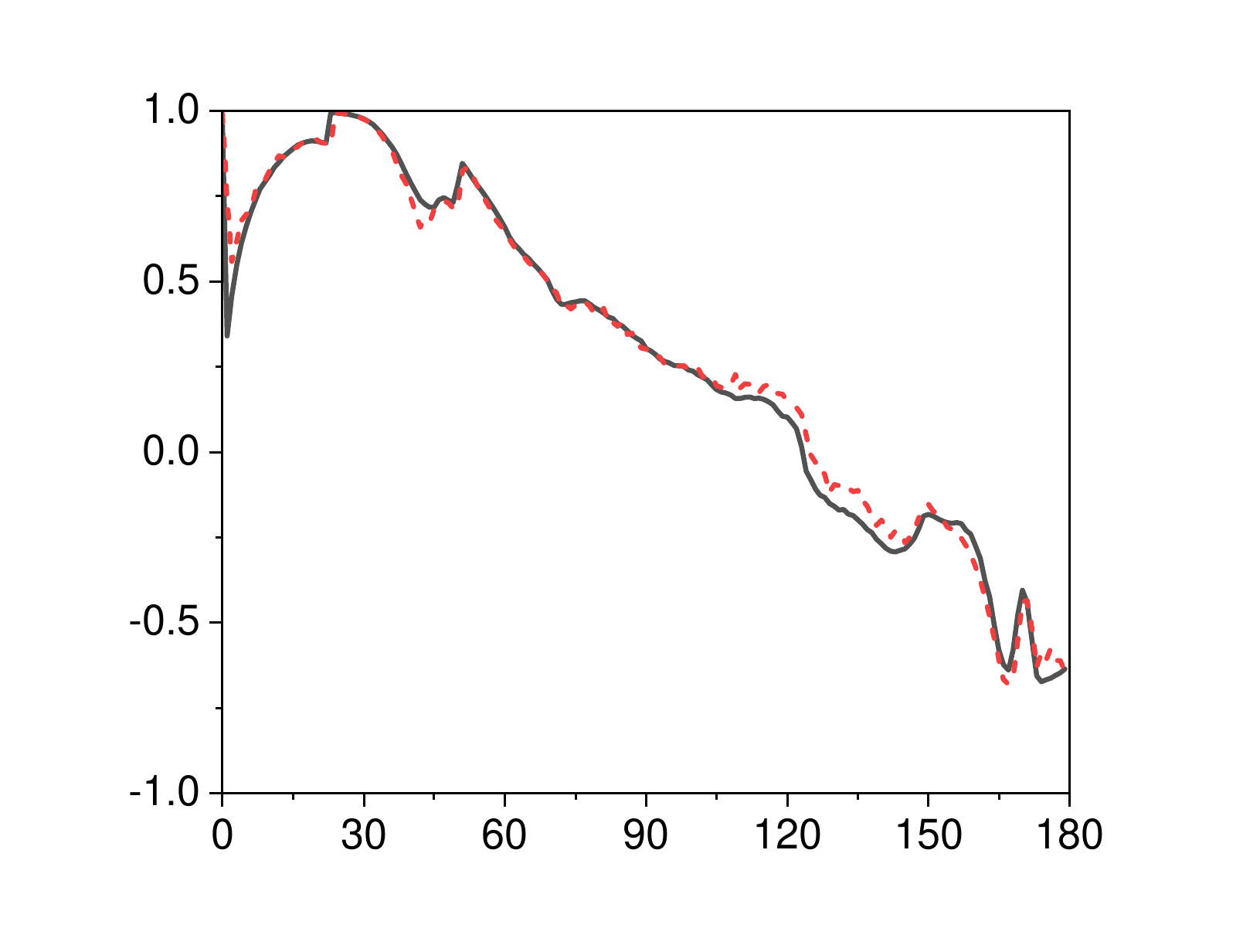}
        \caption{$M_{33}/M_{11}$}
         \label{subFig:5.d}
    \end{subfigure}      
    
     \begin{subfigure}{0.45\textwidth}
        \centering
        \includegraphics[width=\linewidth]{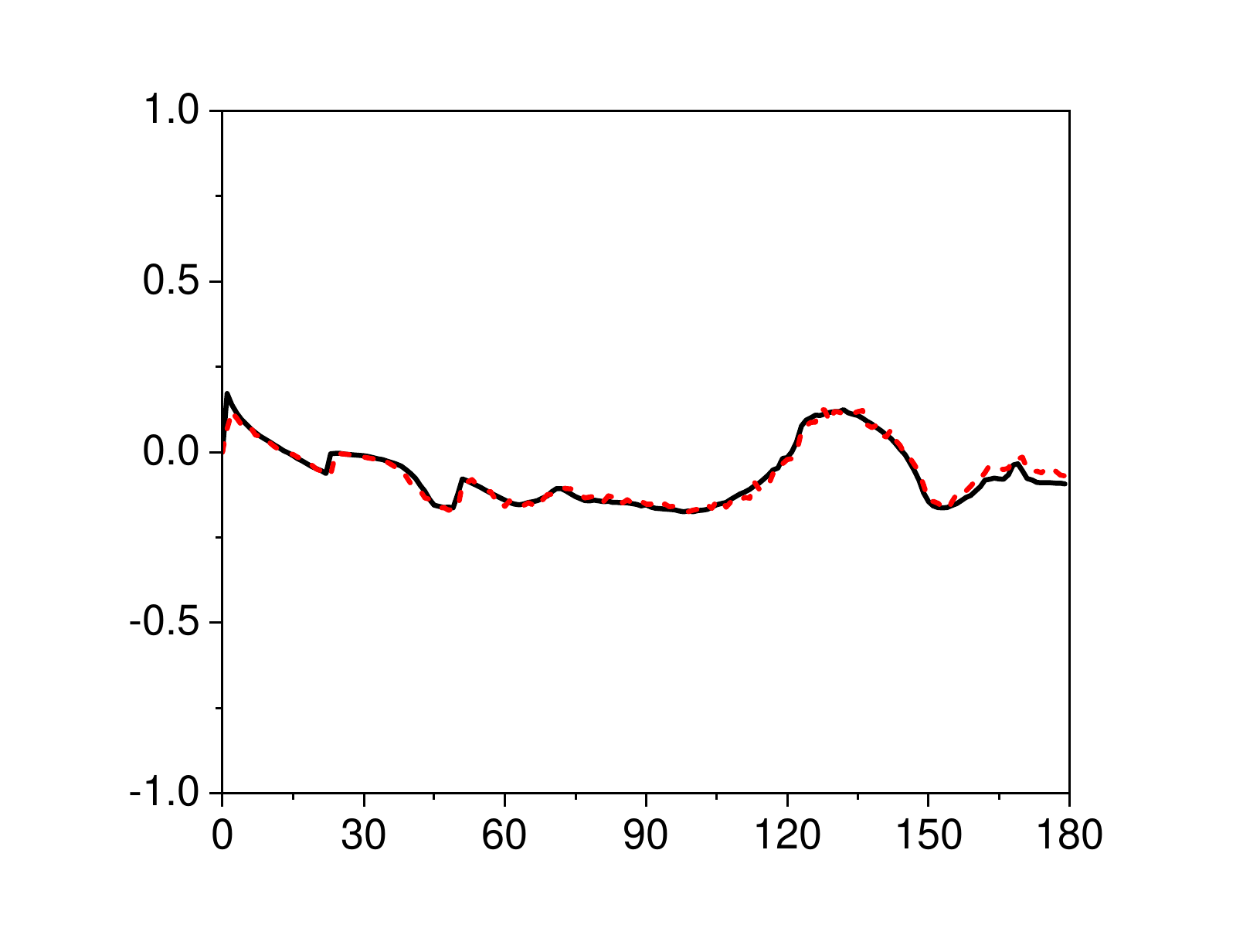}
        \caption{$M_{34}/M_{11}$}
         \label{subFig:5.e}
    \end{subfigure}
    \begin{subfigure}{0.45\textwidth}
        \centering
        \includegraphics[width=\linewidth]{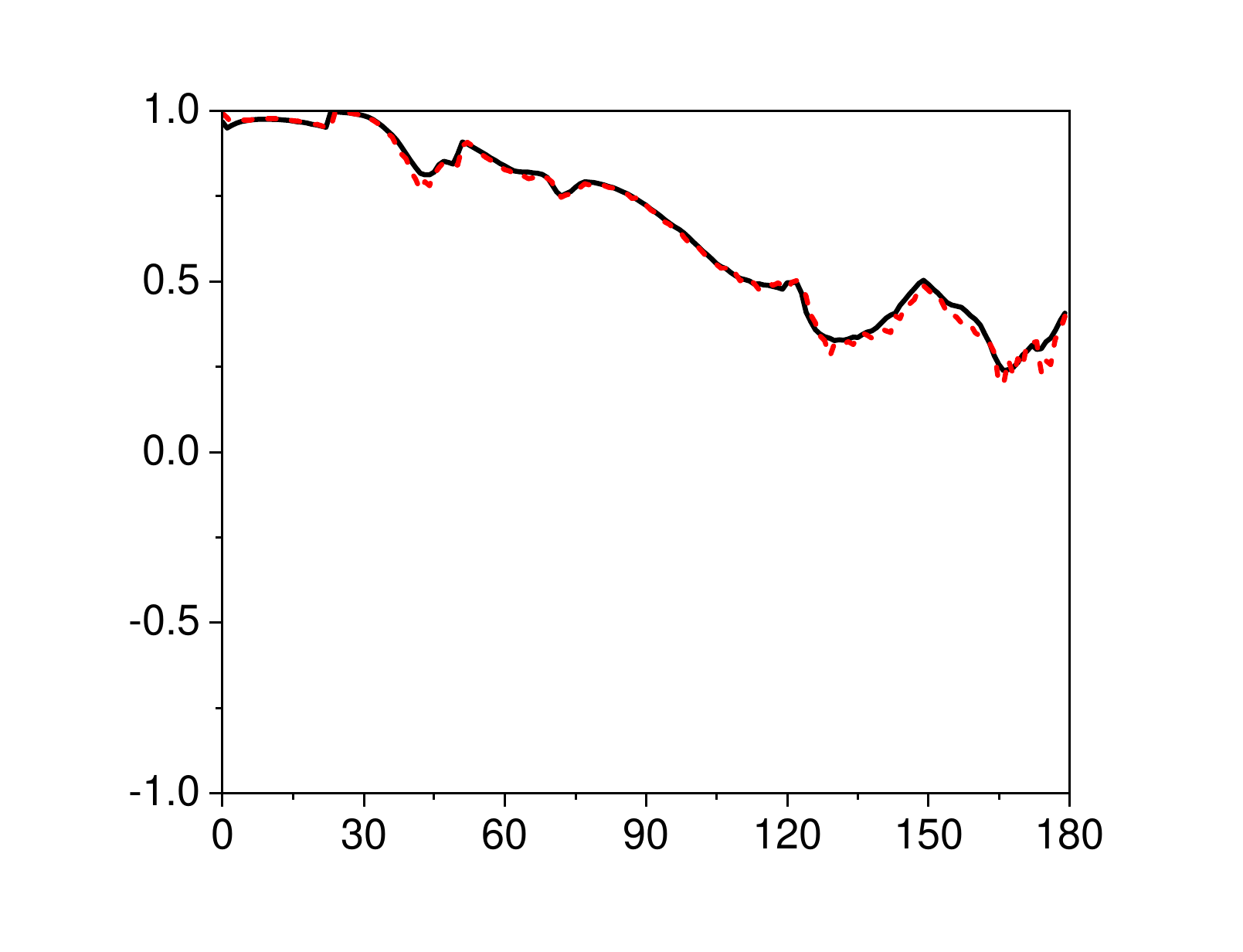}
        \caption{$M_{44}/M_{11}$}
         \label{subFig:5.f}
    \end{subfigure}      
    
    \caption{Comparison of six Mueller matrix elements for randomly oriented hexagonal column obtained by the program \textit{MMCP} (black solid line) and by Macke's method \cite{macke1993scattering, macke1996single, macke_2020_3965488} (red short dash). The horizontal axis represents the scattering angle (in degrees).}
    \label{Fig:5}
\end{figure}

The Mueller matrix computed for randomly oriented hexagonal column is compared with Macke's  results\cite{macke1993scattering, macke1996single, macke_2020_3965488}. The hexagonal column has a height of $200 \,\ \mu\text{m}$ and a base diameter of $80 \,\ \mu\text{m}$. The calculation is performed at a wavelength $0.308 \,\ \mu\text{m}$, the corresponding refractive index of ice is taken as 1.332 and absorption is neglected. The comparison results are presented in Figure \ref{Fig:5}.       

As shown in Figure \ref{Fig:5}, the six Mueller matrix elements for randomly oriented hexagonal column computed using the program \textit{MMCP} developed in this study are in good agreement with those results obtained by Macke’s method. In Figure \ref{Fig:5}(\subref{subFig:5.a}), it can be noticed an offset of the scattering phase function. This shift could be explained by the use of raw data obtained from \cite{macke_2020_3965488}, where diffraction is included and a normalization condition different from Eq. \eqref{Eq:normalization} is applied. For the other five Mueller matrix elements, the two curves are generally consistent, showing only small local differences. Those minor localized discrepancies observed, for example, in Figure \ref{Fig:5}(\subref{subFig:5.b}), \ref{Fig:5}(\subref{subFig:5.c}), and \ref{Fig:5}(\subref{subFig:5.d}), could result from a differences in the sampled number of rays and orientations. In program \textit{MMCP}, the number of traced rays is set to $100$ for each orientation, and the number of sampled orientation is set to $10^6$, whereas the corresponding values in Macke’s calculations are $300$ and $3\times10^4$, respectively. Another possible source of discrepancy is the treatment of total internal reflections. In our scheme, the total number of reflections (including total internal reflections) is limited to 10. In contrast, Macke’s calculation restricts the recursion depth to 10, while total internal reflections are counted separately (with a maximum of 100).

\subsection{Faceted ellipsoid}
\begin{figure}[htbp]
    \centering  
    \begin{subfigure}{0.45\textwidth}
        \centering
        \begin{overpic}[width=\linewidth]{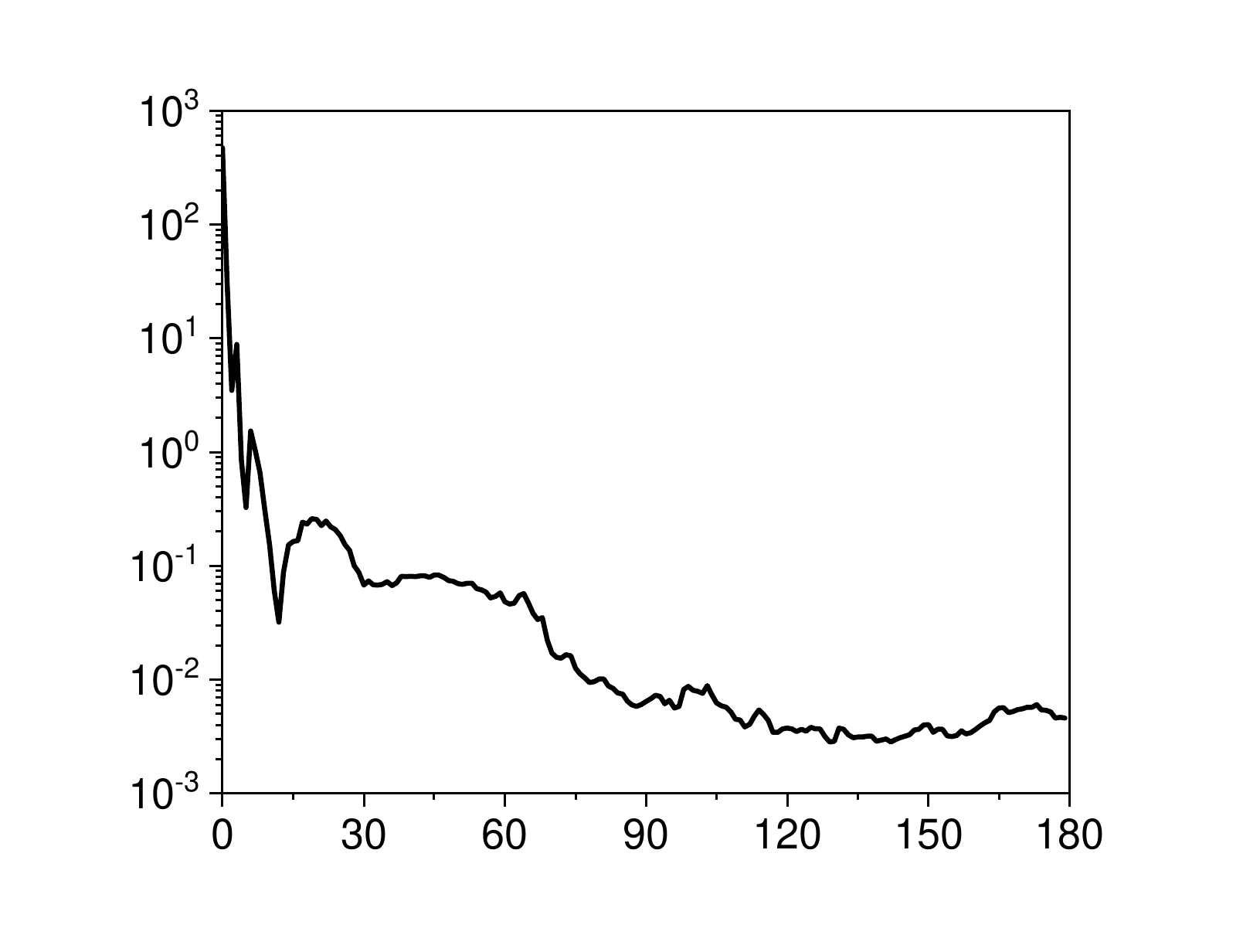}
            \put(35,35){\includegraphics[width=3.5cm]{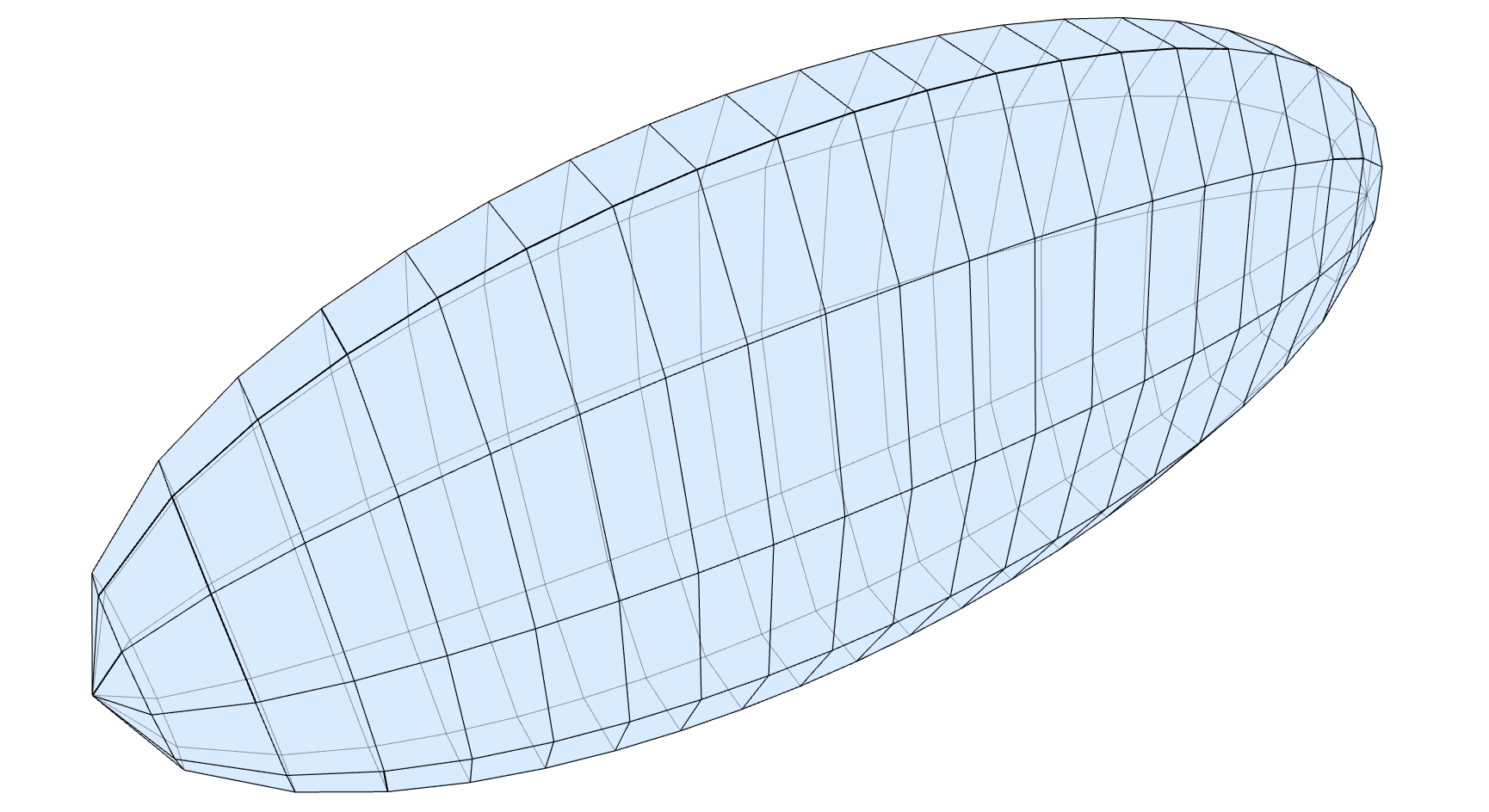}}
        \end{overpic}
        \caption{$M_{11}$}
        \label{subFig:6.a}
    \end{subfigure}
    \begin{subfigure}{0.45\textwidth}
        \centering
        \includegraphics[width=\linewidth]{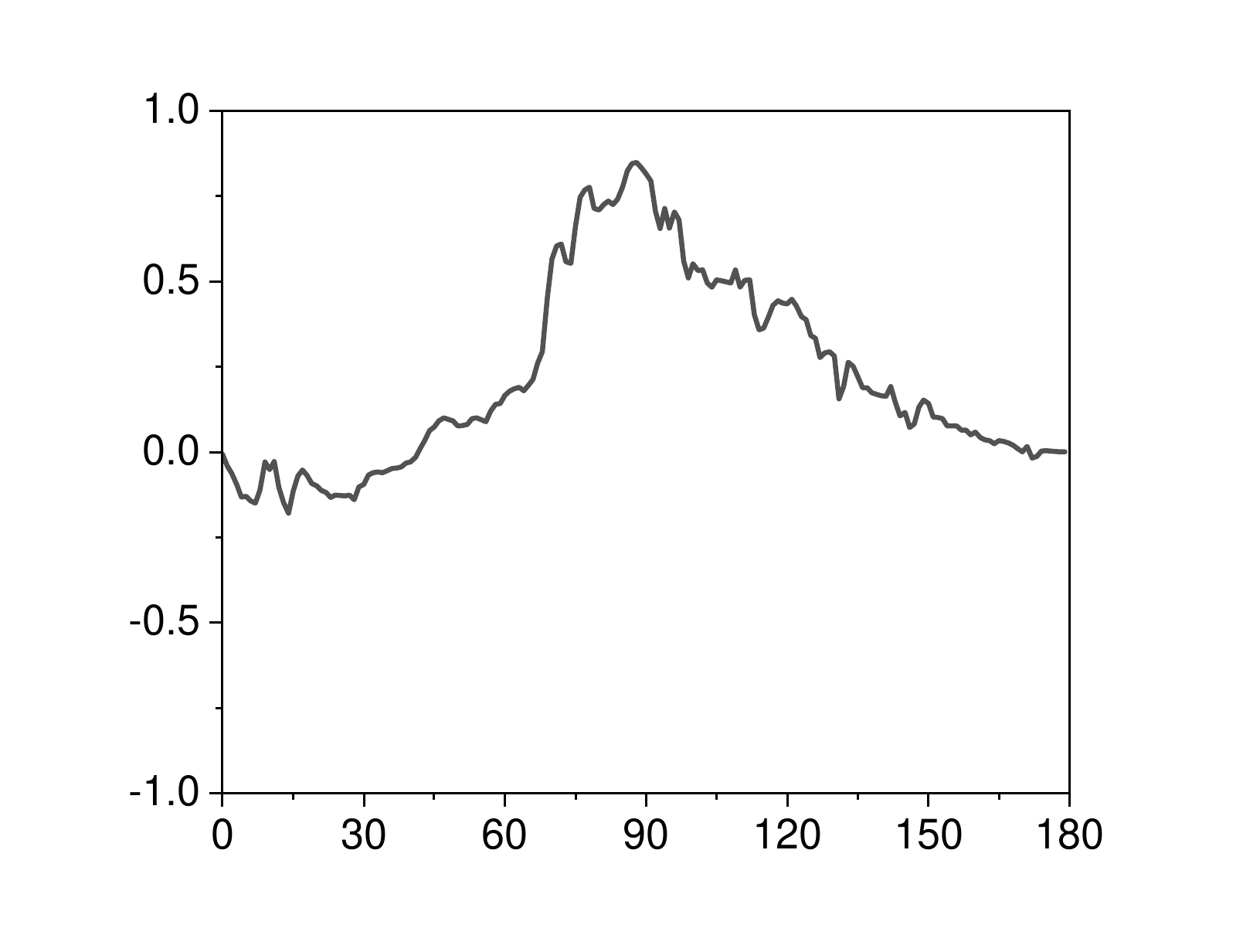}
        \caption{$-M_{12}/M_{11}$}
         \label{subFig:6.b}
    \end{subfigure}      
    
     \begin{subfigure}{0.45\textwidth}
        \centering
        \includegraphics[width=\linewidth]{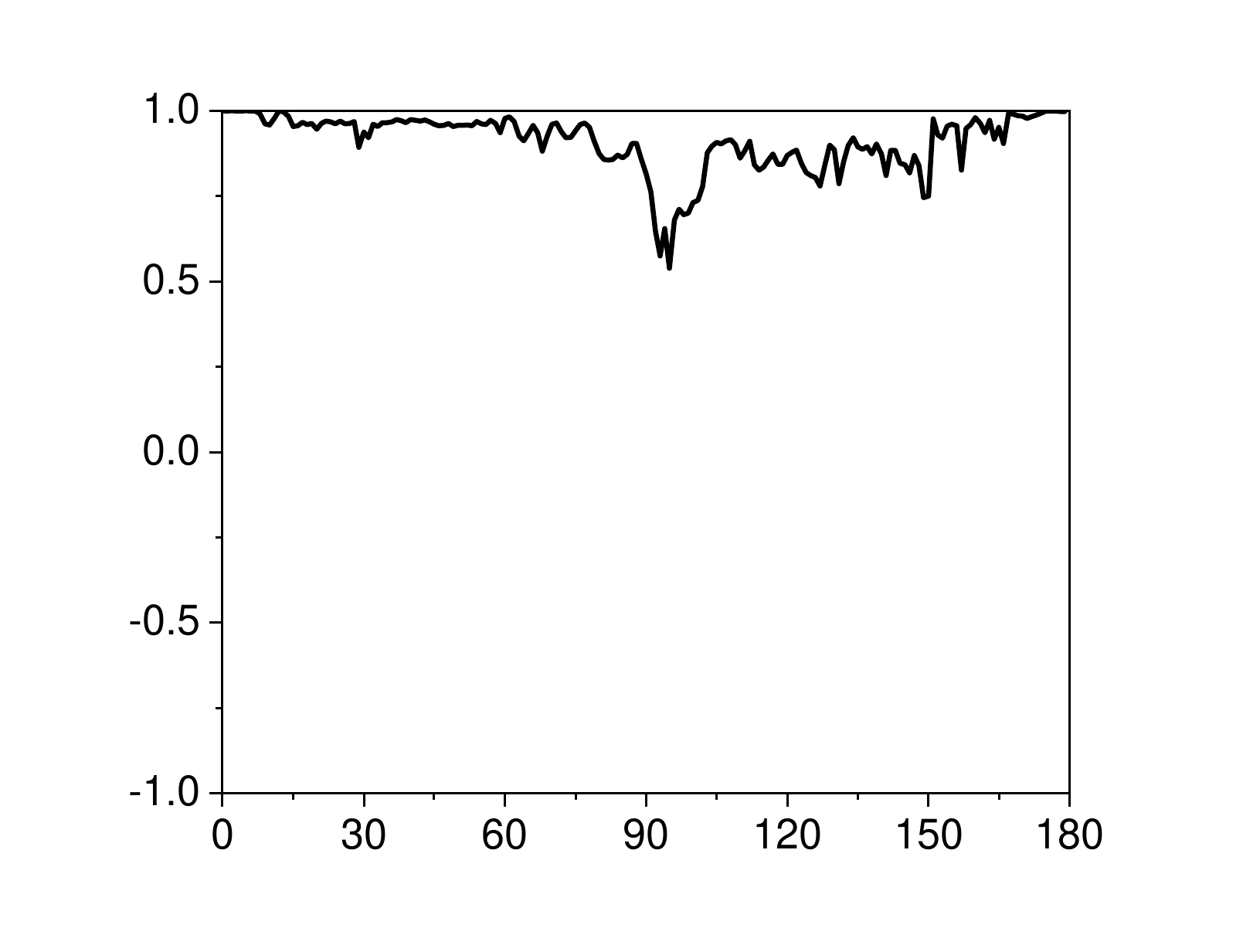}
        \caption{$M_{22}/M_{11}$}
         \label{subFig:6.c}
    \end{subfigure}
    \begin{subfigure}{0.45\textwidth}
        \centering
        \includegraphics[width=\linewidth]{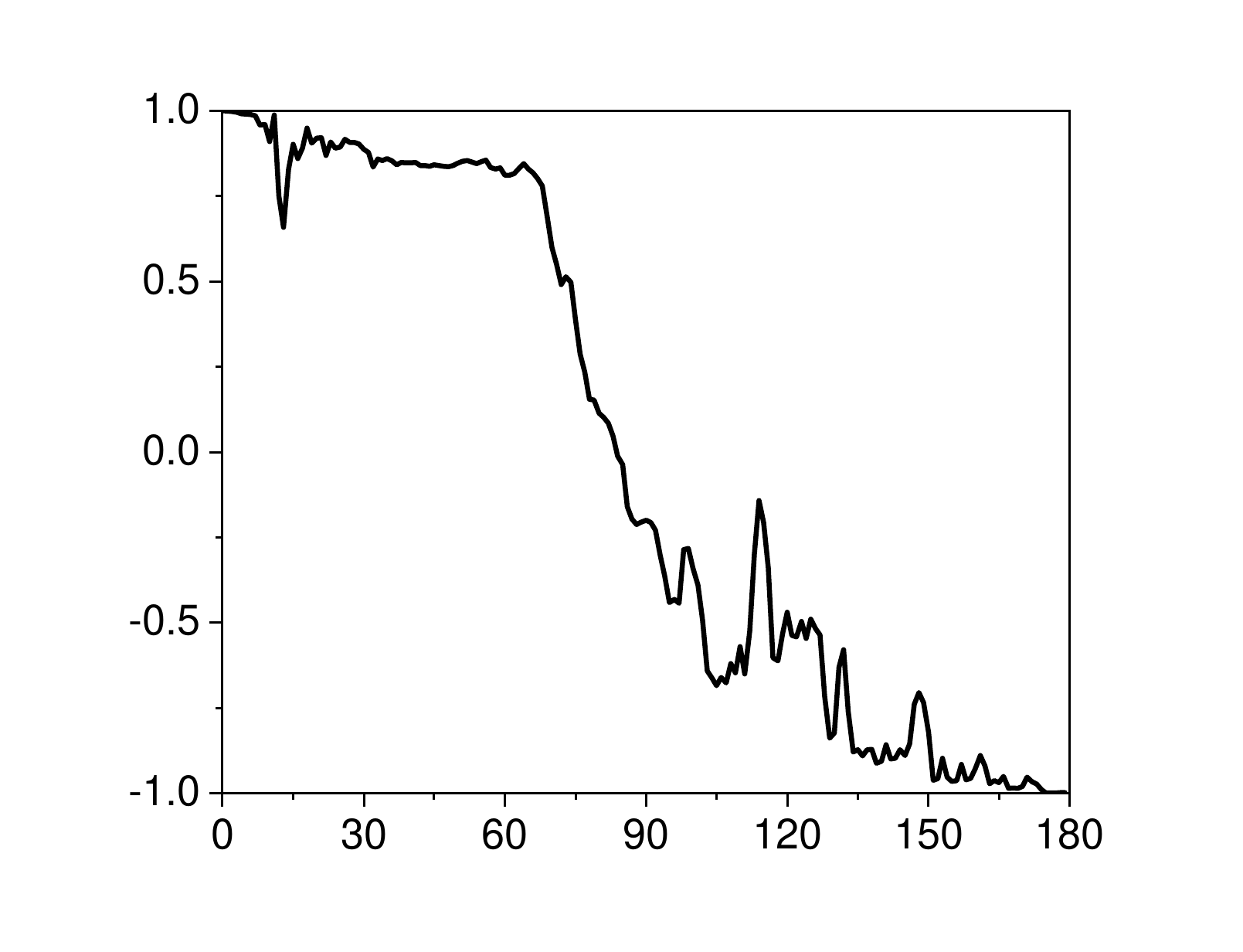}
        \caption{$M_{33}/M_{11}$}
         \label{subFig:6.d}
    \end{subfigure}      
    
     \begin{subfigure}{0.45\textwidth}
        \centering
        \includegraphics[width=\linewidth]{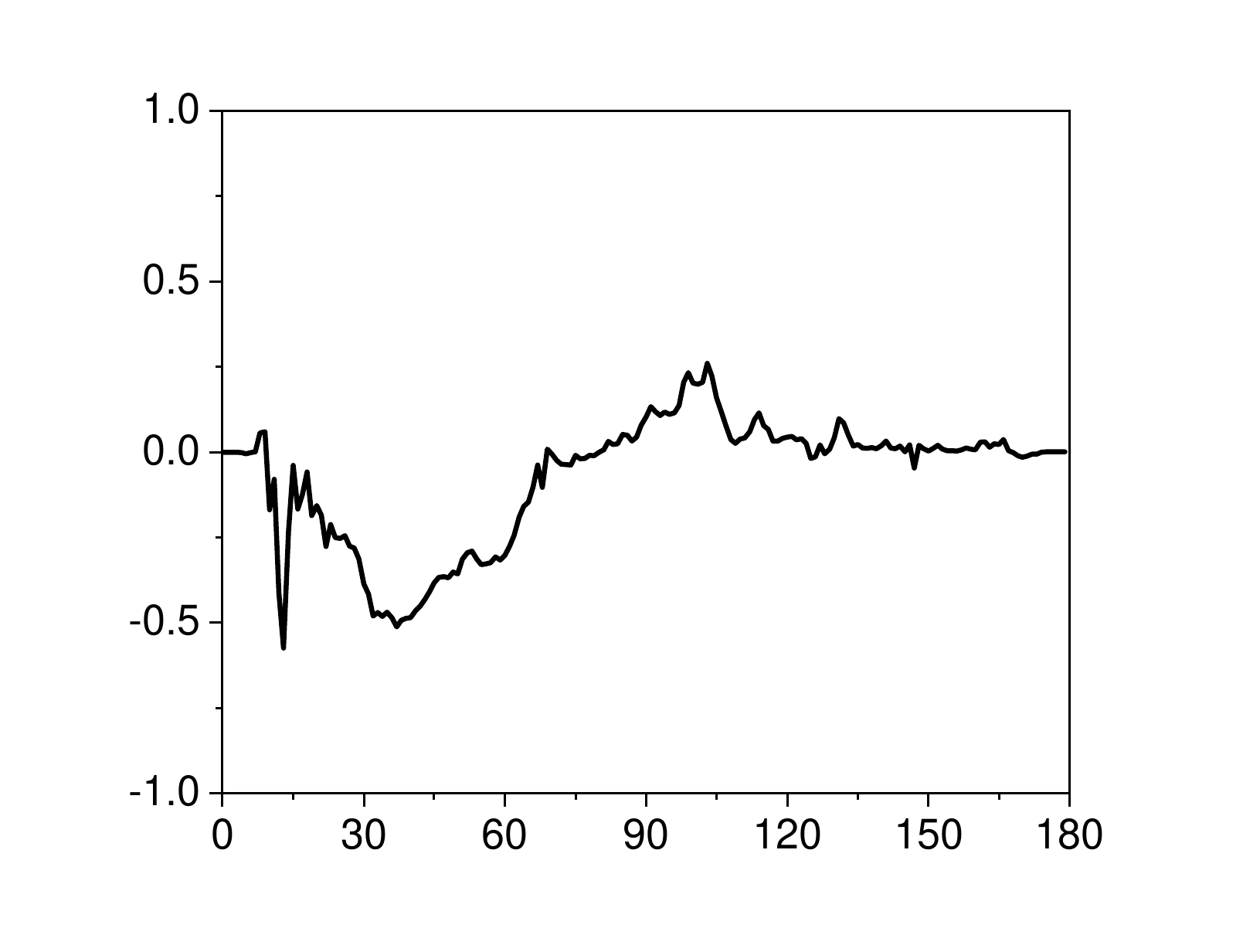}
        \caption{$M_{34}/M_{11}$}
         \label{subFig:6.e}
    \end{subfigure}
    \begin{subfigure}{0.45\textwidth}
        \centering
        \includegraphics[width=\linewidth]{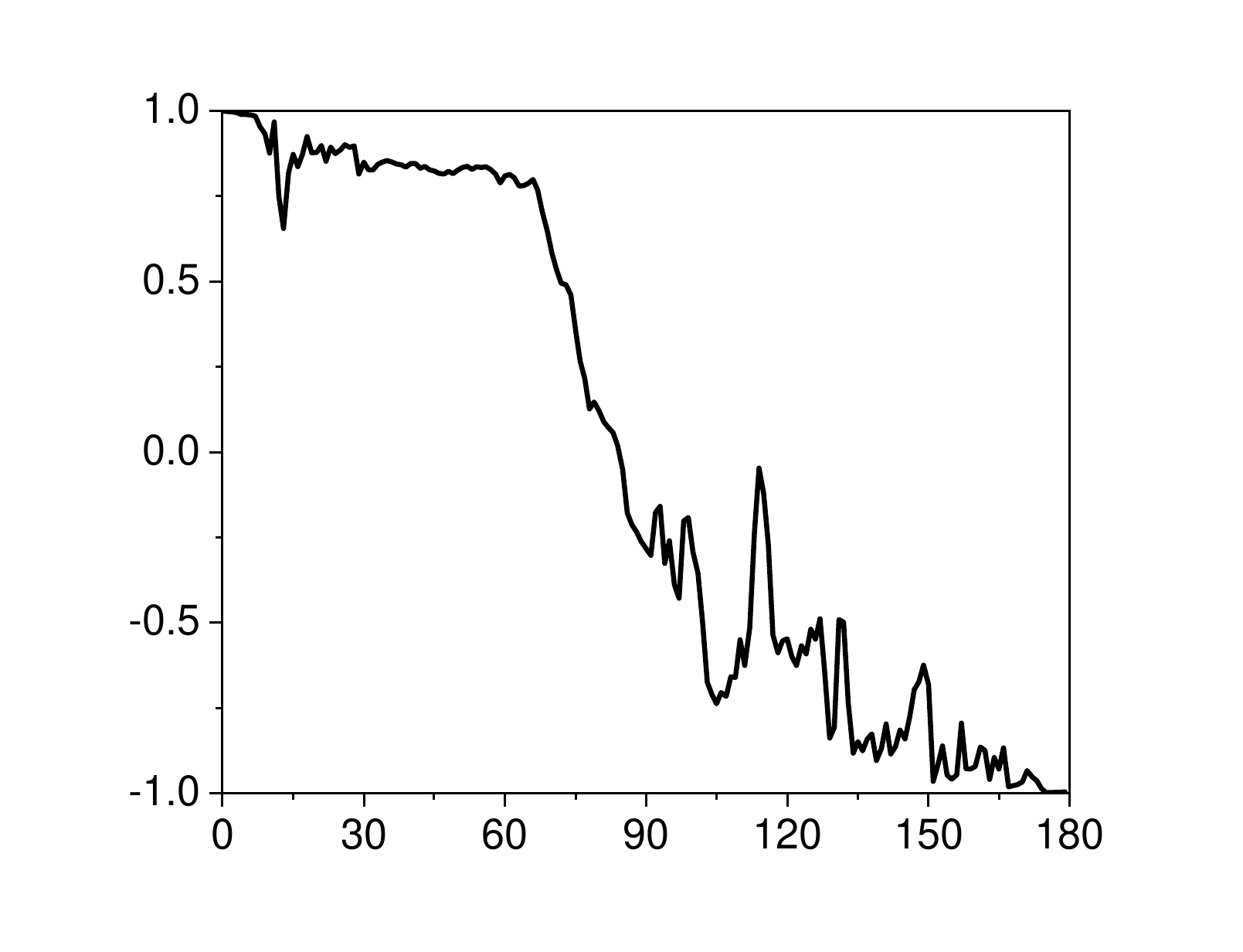}
        \caption{$M_{44}/M_{11}$}
         \label{subFig:6.f}
    \end{subfigure}      
    
    \caption{Mueller matrix elements for faceted ellipsoid obtained by the program \textit{MMCP}. The horizontal axis represents the scattering angle (in degrees).}
    \label{Fig:6}
\end{figure}

Figure \ref{Fig:6} shows the computed scattering matrix of a faceted ellipsoid with semi-axes $a,b,c$ in the ratio $2:5:10$. Initially, all points are defined on the ellipsoidal surface by discretizing the polar angle $\theta$ and the azimuthal angle $\varphi$ in spherical coordinates. A convex hull is then constructed from these points, followed by coplanarity checks and merging of any coplanar faces. The resulting convex polyhedron consists of 202 vertices and 210 faces. A total of 100 rays were traced for each particle orientation, with $10^6$ orientations sampled. The refractive index of ice was set to 1.332, and absorption effects were neglected.

From Figure \ref{Fig:6} it can be noted that $M_{33}/M_{11}$ and $M_{44}/M_{11}$ are very close to each other over the entire scattering angle interval $[0,180^{\circ}]$, and $M_{22}/M_{11}$ is close to 1 except the scattering angle region near $100^{\circ}$. These results suggest that the scattering particles exhibit a certain degree of spherical symmetry, which is consistent with the relations $M_{11}=M_{22}$ and $M_{33}=M_{44}$, as expected for ideal spherical particles.

\subsection{Random convex hull}
\begin{figure}[htbp]
    \centering  
    \begin{subfigure}{0.45\textwidth}
        \centering
        \begin{overpic}[width=\linewidth]{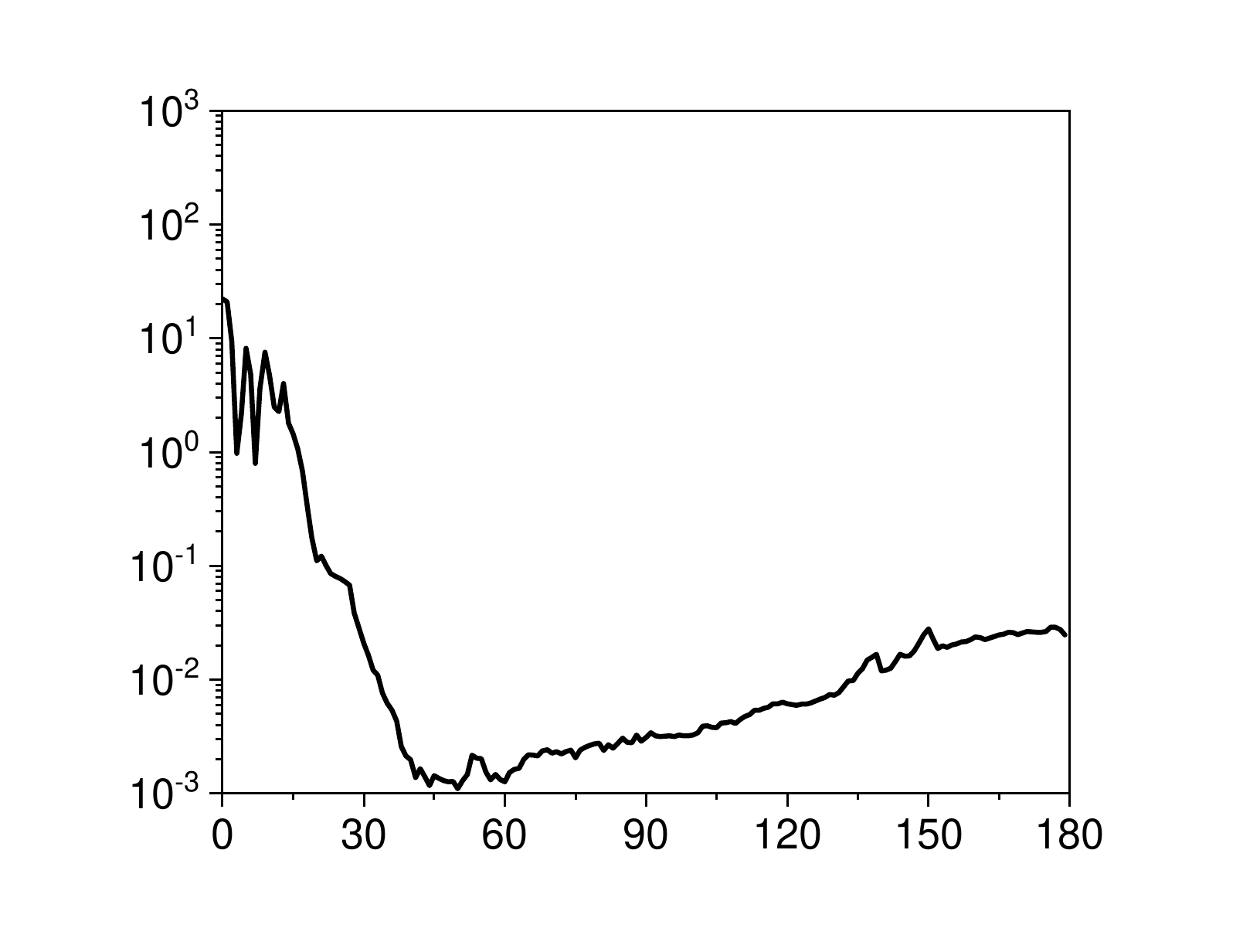}
            \put(39,30){\includegraphics[width=2.3cm]{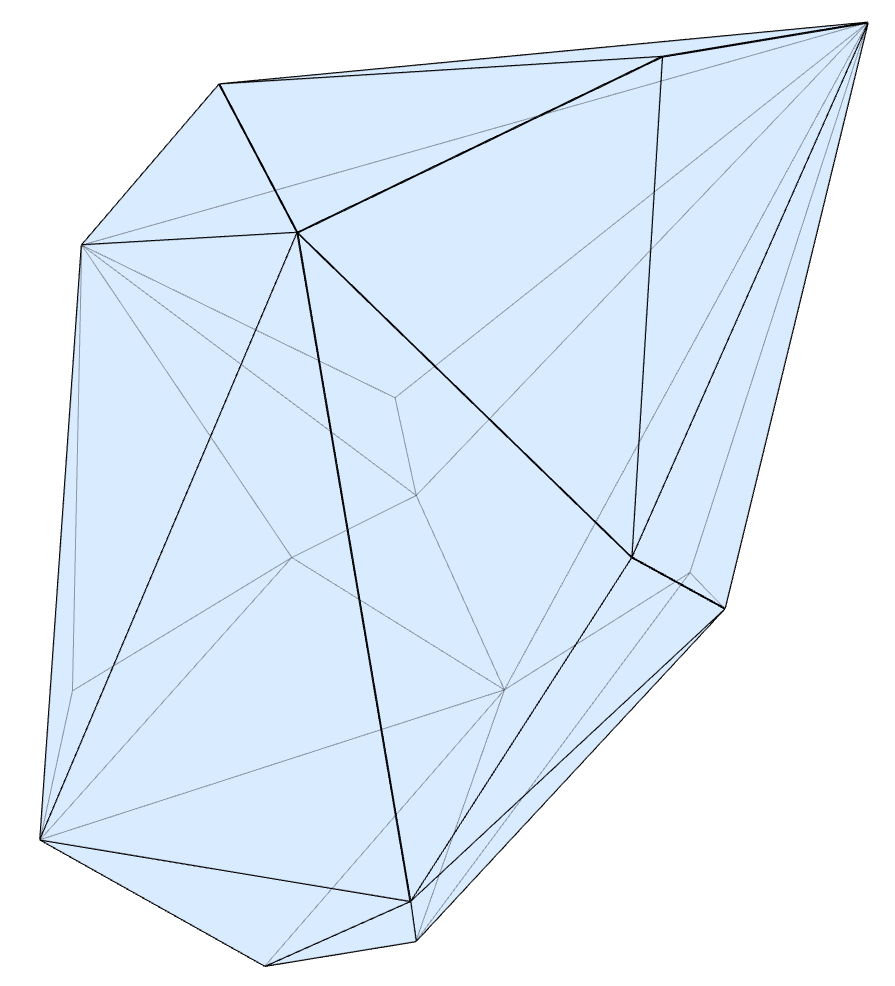}}
        \end{overpic}
        \caption{$M_{11}$}
        \label{subFig:7.a}
    \end{subfigure}
    \begin{subfigure}{0.45\textwidth}
        \centering
        \includegraphics[width=\linewidth]{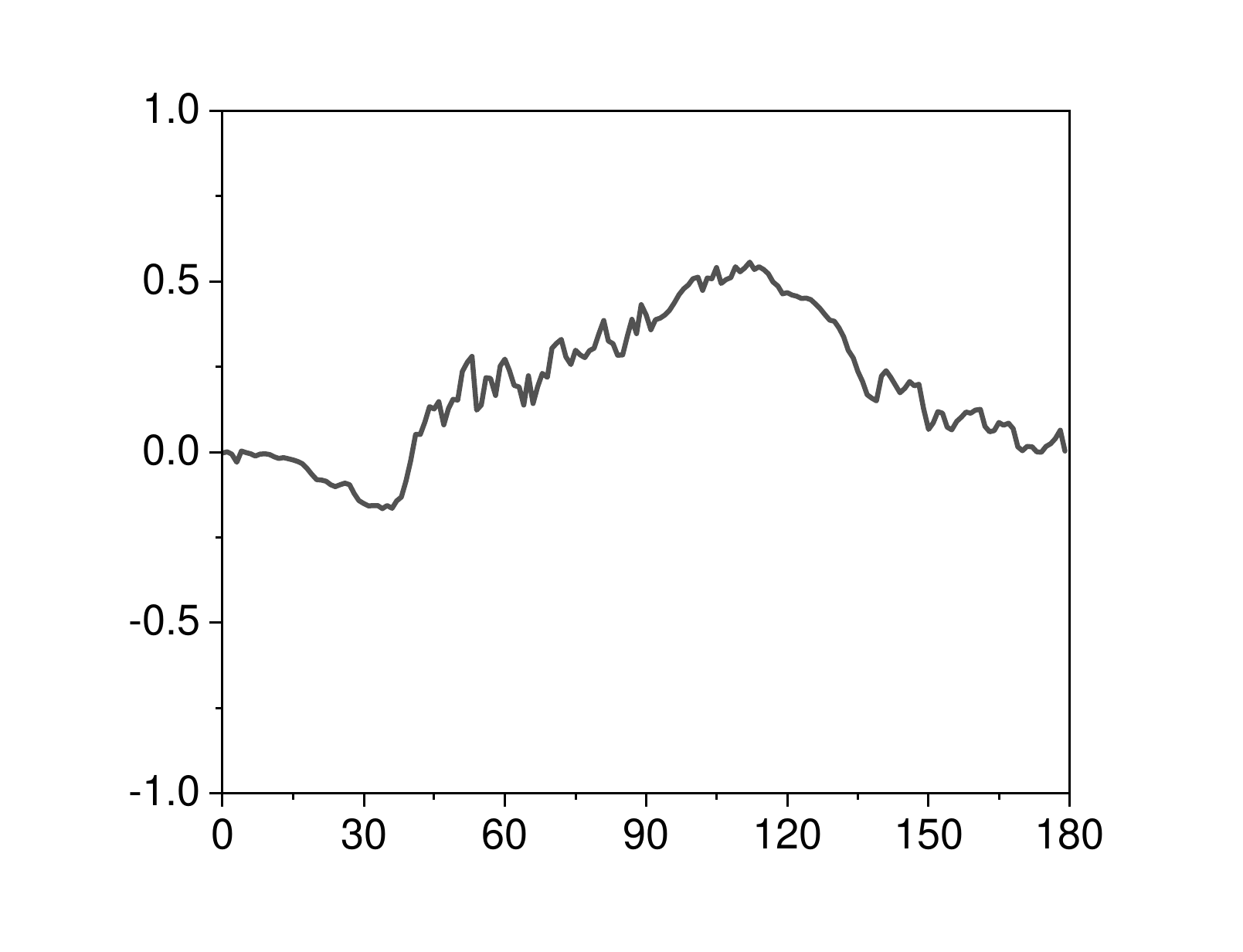}
        \caption{-$M_{12}/M_{11}$}
         \label{subFig:7.b}
    \end{subfigure}      
    
     \begin{subfigure}{0.45\textwidth}
        \centering
        \includegraphics[width=\linewidth]{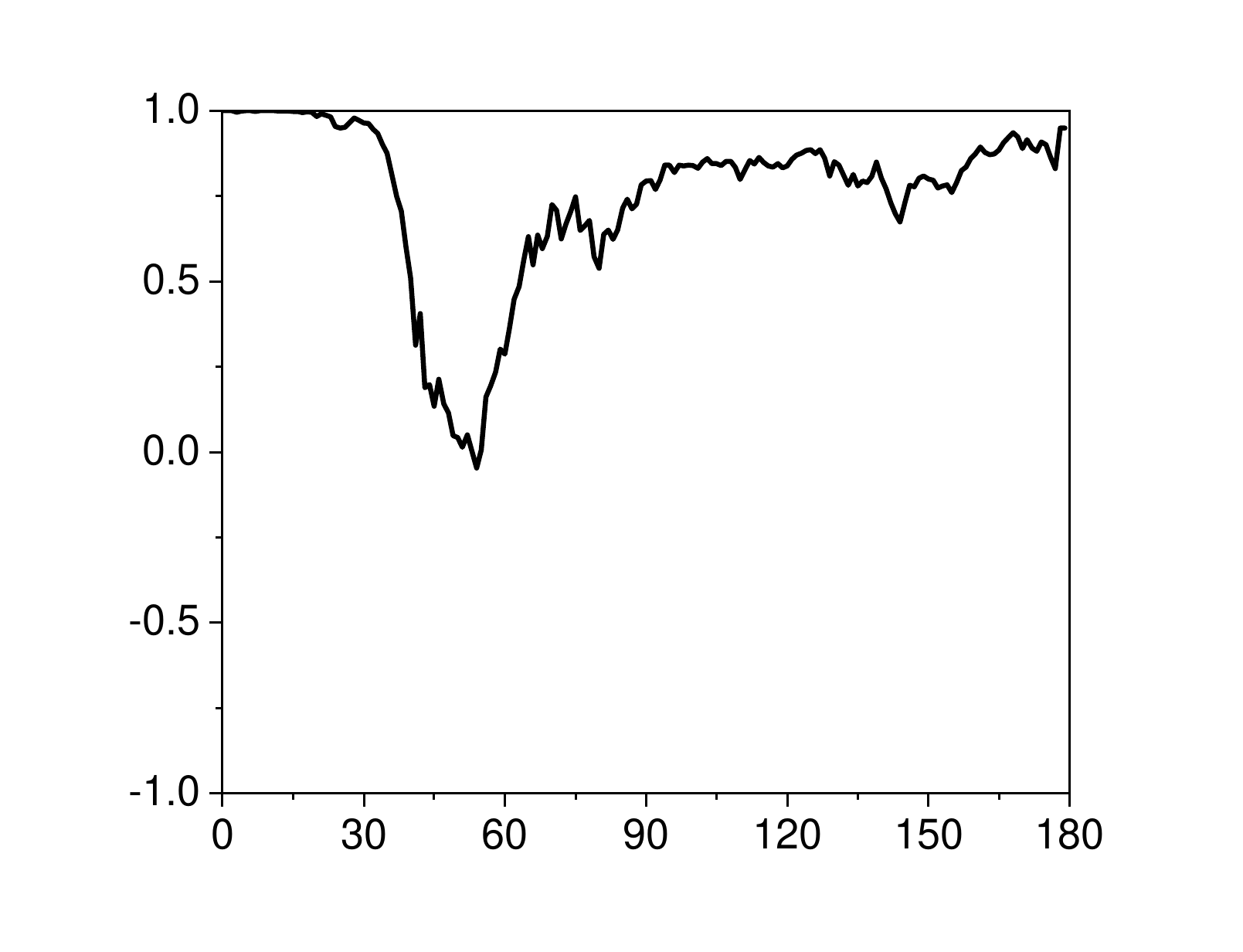}
        \caption{$M_{22}/M_{11}$}
         \label{subFig:7.c}
    \end{subfigure}
    \begin{subfigure}{0.45\textwidth}
        \centering
        \includegraphics[width=\linewidth]{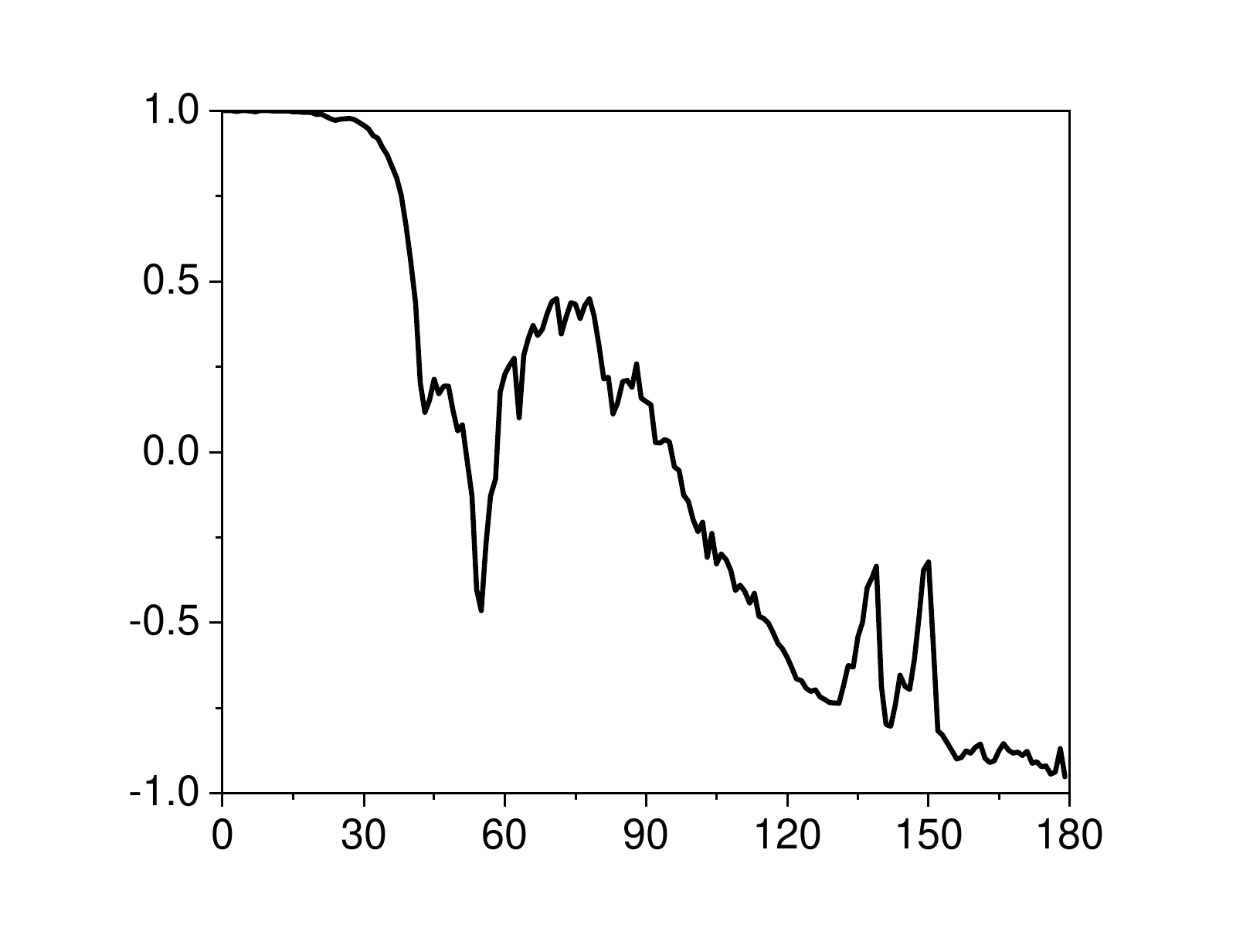}
        \caption{$M_{33}/M_{11}$}
         \label{subFig:7.d}
    \end{subfigure}      
    
     \begin{subfigure}{0.45\textwidth}
        \centering
        \includegraphics[width=\linewidth]{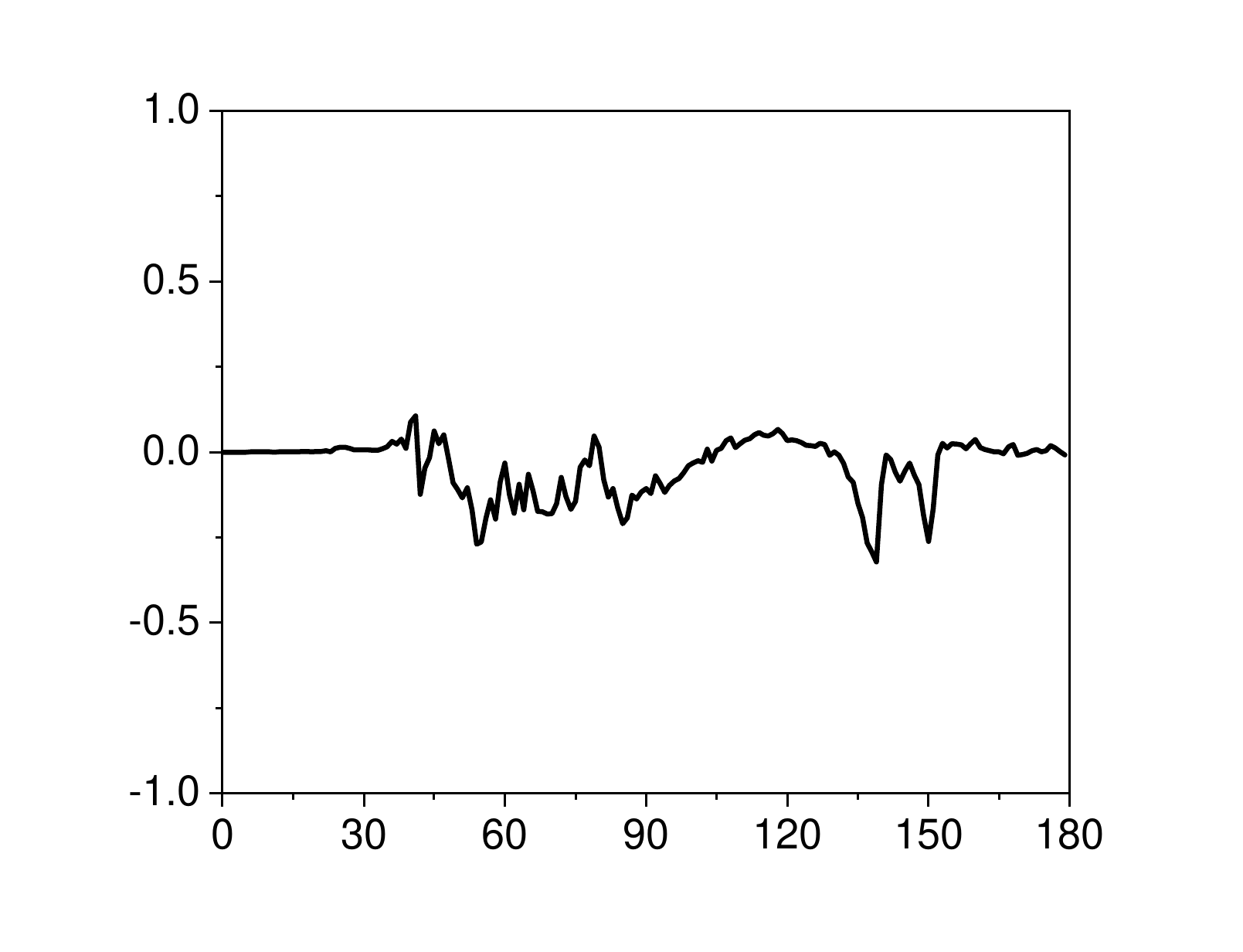}
        \caption{$M_{34}/M_{11}$}
         \label{subFig:7.e}
    \end{subfigure}
    \begin{subfigure}{0.45\textwidth}
        \centering
        \includegraphics[width=\linewidth]{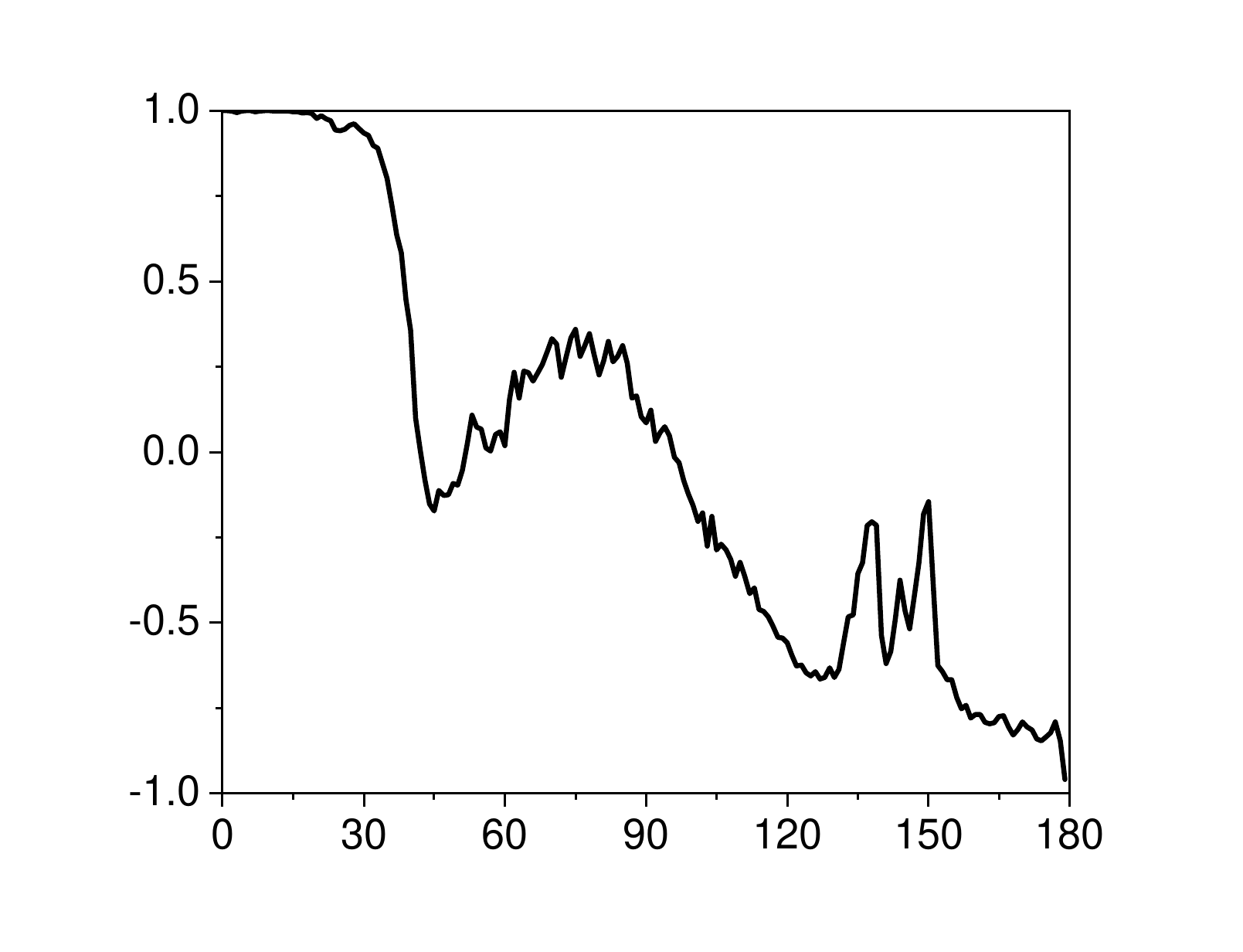}
        \caption{$M_{44}/M_{11}$}
         \label{subFig:7.f}
    \end{subfigure}      
    
    \caption{Mueller matrix elements for randomly constructed convex hull obtained by the program \textit{MMCP}. The horizontal axis represents the scattering angle (in degrees).}
    \label{Fig:7}
\end{figure}

In Figure \ref{Fig:7} the six Mueller matrix elements are presented for a randomly generated convex hull. To generate the convex hull, 25 points are randomly and uniformly sampled within the cube $[-1,1]^3$. The constructed convex hull, shown in Figure \ref{Fig:7} (\subref{subFig:7.a}), comprises 17 vertices and 30 faces. Note that since all the initial points are randomly generated, the probability that four points lie exactly on the same plane is practically zero in a computer system. Therefore, all the faces of the resulting convex hull are triangles. All other parameters, such as the number of ray and orientation, the refractive index, follow the same settings as in the previous computational experiments. It should be restated that, in all computational experiments conducted in this study, absorption and diffraction effects are excluded.  

The absence of a pronounced delta-transmission phenomenon \cite{liou2016light} can be noticed in Figure \ref{Fig:7} (\subref{subFig:7.a}). The delta-transmission near $0^{\circ}$ occurs when the rays are traced for a crystal with parallel or nearly parallel planes. As illustrated in Figure \ref{Fig:7} (\subref{subFig:7.a}), the constructed convex hull contains no parallel planes, and therefore, the characteristic delta forward peak does not appear. This conclusion is further supported by a comparison of the first scattering matrix element $M_{11}$ across Figures \ref{Fig:5}, \ref{Fig:6}, and \ref{Fig:7}. For the newly tested particle model as shown in Figures \ref{Fig:6} and \ref{Fig:7}, the correctness of the the scattering matrix calculations can be preliminarily assessed using certain specific relations, such as for the scattering angles 0 and $\pi$ \cite{mishchenko2000light,hulst1981light}: $M_{22}(0)=M_{33}(0), M_{22}(\pi)=-M_{33}(\pi), M_{12}(0)=M_{34}(0)=M_{12}(\pi)=M_{34}(\pi)=0$. As shown by the computational results in Figures \ref{Fig:6} and \ref{Fig:7}, these relations are satisfied.

\section{Conclusions}
\label{section 4}
A unified scattering matrix computational framework is developed on the basis of the convex hull algorithm and the ray tracing principle. The proposed approach for model construction and computation of light scattering matrices is universally applicable to convex polyhedral particles in the geometrical optics regime. Absorption and diffraction effects are excluded in this study. The computational results show that the six Mueller matrix elements of randomly oriented hexagonal columns obtained in this study generally agree well with those calculated by Macke \cite{macke1993scattering, macke1996single, macke_2020_3965488}. From the computed scattering matrix elements shown in Figures \ref{Fig:5}, \ref{Fig:6}, and \ref{Fig:7}, it is evident that the scattering and polarization characteristics are highly sensitive to the details of particle geometry. 

In reality, ice crystals are often more complicated than the convex polyhedra considered here; for instance, they can be concave or form aggregates. Nevertheless, the framework presented here, along with the implemented C++ code \textit{MMCP}, offers an efficient tool for simulating light scattering by ice crystals or other convex particles of arbitrary shape, and may therefore prove valuable for studies of atmospheric radiative transfer and related optical modeling. In addition, the framework can be further improved by incorporating diffraction and absorption, and it can be further extended to the study of multiple scattering and oriented particles in optically anisotropic ice clouds.

\bibliographystyle{unsrt}
\bibliography{2025Paper01}

\begin{thebibliography}{10}

\bibitem{mishchenko2000light}
Michael~I Mishchenko, Joop~W Hovenier, and Larry~D Travis.
\newblock Light scattering by nonspherical particles: theory, measurements, and applications.
\newblock {\em Measurement Science and Technology}, 11(12):1827--1827, 2000.

\bibitem{liou2016light}
Kuo-Nan Liou and Ping Yang.
\newblock {\em Light scattering by ice crystals: fundamentals and applications}.
\newblock Cambridge University Press, 2016.

\bibitem{kokhanovsky1998dependence}
AA~Kokhanovsky and TY~Nakajima.
\newblock The dependence of phase functions of large transparent particles on their refractive index and shape.
\newblock {\em Journal of Physics D: Applied Physics}, 31(11):1329, 1998.

\bibitem{macke1996applicability}
Andreas Macke and Michael~I Mishchenko.
\newblock Applicability of regular particle shapes in light scattering calculations for atmospheric ice particles.
\newblock {\em Applied optics}, 35(21):4291--4296, 1996.

\bibitem{takano1995radiative}
Y~Takano and KN~Liou.
\newblock Radiative transfer in cirrus clouds. part iii: Light scattering by irregular ice crystals.
\newblock {\em Journal of Atmospheric Sciences}, 52(7):818--837, 1995.

\bibitem{GRYNKO2003319}
Ye. Grynko and Yu. Shkuratov.
\newblock Scattering matrix calculated in geometric optics approximation for semitransparent particles faceted with various shapes.
\newblock {\em Journal of Quantitative Spectroscopy and Radiative Transfer}, 78(3):319--340, 2003.

\bibitem{murray2015trigonal}
Benjamin~J Murray, Christoph~G Salzmann, Andrew~J Heymsfield, Steven Dobbie, Ryan~R Neely~III, and Christopher~J Cox.
\newblock Trigonal ice crystals in earth’s atmosphere.
\newblock {\em Bulletin of the American Meteorological Society}, 96(9):1519--1531, 2015.

\bibitem{yang2018review}
Ping Yang, Souichiro Hioki, Masanori Saito, Chia-Pang Kuo, Bryan~A Baum, and Kuo-Nan Liou.
\newblock A review of ice cloud optical property models for passive satellite remote sensing.
\newblock {\em Atmosphere}, 9(12):499, 2018.

\bibitem{liu2014effective}
Chao Liu, R~Lee Panetta, and Ping Yang.
\newblock The effective equivalence of geometric irregularity and surface roughness in determining particle single-scattering properties.
\newblock {\em Optics express}, 22(19):23620--23627, 2014.

\bibitem{macke1996single}
Andreas Macke, Johannes Mueller, and Ehrhard Raschke.
\newblock Single scattering properties of atmospheric ice crystals.
\newblock {\em Journal of Atmospheric Sciences}, 53(19):2813--2825, 1996.

\bibitem{liu2013effects}
Chao Liu, R~Lee Panetta, and Ping Yang.
\newblock The effects of surface roughness on the scattering properties of hexagonal columns with sizes from the rayleigh to the geometric optics regimes.
\newblock {\em Journal of Quantitative Spectroscopy and Radiative Transfer}, 129:169--185, 2013.

\bibitem{gasteiger2011modelling}
Josef Gasteiger, Matthias Wiegner, Silke Gro{\ss}, Volker Freudenthaler, Carlos Toledano, Matthias Tesche, and Konrad Kandler.
\newblock Modelling lidar-relevant optical properties of complex mineral dust aerosols.
\newblock {\em Tellus B: Chemical and Physical Meteorology}, 63(4):725--741, 2011.

\bibitem{shishko2019coherent}
Victor Shishko, Alexander Konoshonkin, Natalia Kustova, Dmitry Timofeev, and Anatoli Borovoi.
\newblock Coherent and incoherent backscattering by a single large particle of irregular shape.
\newblock {\em Optics Express}, 27(23):32984--32993, 2019.

\bibitem{mu2022computer}
Q~Mu, BA~Kargin, and EG~Kablukova.
\newblock Computer-aided construction of three-dimensional convex bodies of arbitrary shapes.
\newblock {\em Computational technologies}, 27(2):54--61, 2022.

\bibitem{kargin2022numerical}
BA~Kargin, EG~Kablukova, and Q~Mu.
\newblock Numerical stochastic simulation of optical radiation scattering by ice crystals of irregular random shapes.
\newblock {\em Computational Technologies}, 27(2):4--18, 2022.

\bibitem{preparata2012computational}
Franco~P Preparata and Michael~I Shamos.
\newblock {\em Computational geometry: an introduction}.
\newblock Springer Science \& Business Media, 2012.

\bibitem{de2008computational}
Mark De~Berg, Otfried Cheong, Marc Van~Kreveld, and Mark Overmars.
\newblock {\em Computational geometry: algorithms and applications}.
\newblock Springer, 2008.

\bibitem{kargin2024monte}
BA~Kargin, EG~Kablukova, Q~Mu, and SM~Prigarin.
\newblock Monte carlo method for numerical simulation of solar energy radiation transfer in crystal clouds.
\newblock {\em Numerical Analysis and Applications}, 17(2):140--151, 2024.

\bibitem{Zhang:2004}
Zhibo Zhang, Ping Yang, George~W. Kattawar, Si-Chee Tsay, Bryan~A. Baum, Yongxiang Hu, Andrew~J. Heymsfield, and Jens Reichardt.
\newblock Geometrical-optics solution to light scattering by droxtal ice crystals.
\newblock {\em Appl. Opt.}, 43(12):2490--2499, Apr 2004.

\bibitem{yang2006light}
Ping Yang and Kuo-Nan Liou.
\newblock Light scattering and absorption by nonspherical ice crystals.
\newblock In {\em Light Scattering Reviews: Single and Multiple Light Scattering}, pages 31--71. Springer, 2006.

\bibitem{born2019principles}
M~Born, E~Wolf, and P~Knight.
\newblock Principles of optics (60th anniversary of first edition, 20th anniversary of seventh edition), 2019.

\bibitem{hulst1981light}
Hendrik~Christoffel Hulst and Hendrik~C van~de Hulst.
\newblock {\em Light scattering by small particles}.
\newblock Courier Corporation, 1981.

\bibitem{hovenier2014transfer}
Joop~W Hovenier, Cornelis~VM Van~der Mee, and Helmut Domke.
\newblock {\em Transfer of polarized light in planetary atmospheres: basic concepts and practical methods}, volume 318.
\newblock Springer Science \& Business Media, 2014.

\bibitem{fujiwara2007spectroscopic}
Hiroyuki Fujiwara.
\newblock {\em Spectroscopic ellipsometry: principles and applications}.
\newblock John Wiley \& Sons, 2007.

\bibitem{konoshonkin2015beam01}
AV~Konoshonkin, NV~Kustova, and AG~Borovoi.
\newblock Beam splitting algorithm for the problem of light scattering by atmospheric ice crystals. part 1. theoretical foundations of the algorithm.
\newblock {\em Atmospheric and Oceanic Optics}, 28(5):441--447, 2015.

\bibitem{konoshonkin2015beam02}
AV~Konoshonkin, NV~Kustova, and AG~Borovoi.
\newblock Beam splitting algorithm for the problem of light scattering by atmospheric ice crystals. part 2. comparison with the ray tracing algorithm.
\newblock {\em Atmospheric and Oceanic Optics}, 28:448--454, 2015.

\bibitem{macke1993scattering}
Andreas Macke.
\newblock Scattering of light by polyhedral ice crystals.
\newblock {\em Applied Optics}, 32(15):2780--2788, 1993.

\bibitem{macke_2020_3965488}
A.~Macke.
\newblock rt-crystal, 2020.
\newblock Available at \url{https://doi.org/10.5281/zenodo.3965488}.

\end{thebibliography}

\end{sloppypar}
\end{document}